\documentclass[12pt]{article}
\newcommand{\blind}{0}
\usepackage{amsmath,amsfonts,amsthm,amssymb,bm,bbm}
\usepackage{graphicx}
\graphicspath{{figures/}}
\usepackage{enumerate,natbib,caption,placeins,multirow,booktabs,array,tabularx,url,color}
\usepackage{algorithm,algorithmic}
\usepackage[colorlinks,citecolor=blue,urlcolor=blue]{hyperref}

\DeclareMathOperator*{\argmin}{arg\,min}

\newcommand{\E}{\mathbb E}
\newcommand{\F}{\mathcal F}
\newcommand{\norm}[1]{\left\lVert#1\right\rVert}

\newcommand{\floor}[1]{\left\lfloor#1\right\rfloor}

\newtheorem{theorem}{Theorem}

\newtheorem{proposition}{Proposition}
\newtheorem{corollary}{Corollary}
\newtheorem{condition}{Condition}
\newtheorem{remark}{Remark}

\begin{document}
\def\spacingset#1{\renewcommand{\baselinestretch}{#1}\small\normalsize}
\spacingset{1}
\renewcommand{\theHequation}{\thesection.\arabic{equation}}
\ifnum\blind=0
\title{\bf Online augmented plug-in inference for multi-group fair prediction}
\author{Enze Shi\textsuperscript{1}, Jinhan Xie\textsuperscript{2},
Linglong Kong\textsuperscript{1}, and Bei Jiang\textsuperscript{1}\thanks{Corresponding author. Email: \url{bei1@ualberta.ca}.}}
\date{}
\maketitle
\vspace{-1cm}
\begin{center}
\textsuperscript{1}Department of Mathematical and Statistical Sciences, University of Alberta\\
\textsuperscript{2}Yunnan Key Laboratory of Statistical Modeling and Data Analysis, Yunnan University
\end{center}
\else
\bigskip\bigskip\bigskip
\begin{center}{\LARGE\bf Online augmented plug-in inference for multi-group fair prediction}\end{center}
\medskip
\fi

\begin{abstract}
Fairness regularization changes both a prediction rule and its sampling
uncertainty. We develop online inference for classifiers trained with a
covariance penalty on group mean decision scores, a tractable proxy for
demographic parity. The framework includes logistic regression and smoothed
support-vector machines, with linear regression as a special case targeting
equality of group mean predictions. The inferential challenge arises because
the fairness operator is estimated from the same stream used to update the
predictor. Each observation contributes to its current loss gradient and to
subsequent penalty estimates, creating an additional first-order fluctuation
in the Polyak--Ruppert average. We derive an asymptotic expansion with harmonic
reuse weights and an augmented covariance incorporating score variation,
operator variation, and their dependence. Exact fixed-memory recursions yield
a consistent covariance estimator and asymptotically valid confidence
intervals for the penalized coefficients. Simulation studies assess coverage
and interval contraction. Adult-income classification and IPUMS wage
regression illustrate coefficient uncertainty and fairness--utility
trade-offs, with classification demographic parity evaluated separately
through threshold decisions.
\end{abstract}

\vfill\newpage
\spacingset{1.9}

\section{Introduction}\label{sec:introduction}
\renewcommand{\theequation}{1.\arabic{equation}}\setcounter{equation}{0}

Prediction systems shape access to credit, employment, public services,
health care, and criminal justice
\citep{potash2015predictive,fuster2022predictably}. Their statistical quality
depends on both predictive performance and the distribution of outputs across groups.
Individual fairness, demographic parity, equalized odds, and related criteria
formalize different normative goals
\citep{dwork2012fairness,hardt2016equality,barocas2023fairness,
caton2024fairness}. Our primary application is demographic parity for a threshold
classifier, which requires equal positive-decision probabilities across
groups. Fitting uses a convex penalty on group mean decision scores as a
proxy; its reduction need not reduce the disparity in threshold decisions.
Our inference concerns the resulting penalized coefficients. For linear
regression, score parity is equality of group mean predictions.

Fairness interventions may act before, during, or after model fitting.
Dependence regularizers and decision-boundary covariance constraints provide
tractable in-processing rules
\citep{kamishima2012fairness,zafar2019fairness}. Reduction-based methods
handle explicit rate or group-loss constraints
\citep{agarwal2019fair}, while optimized score transformations impose linear
restrictions on conditional score means \citep{wei2021optimized}.
Representation and adversarial methods instead modify the learned features or
their distribution \citep{zhang2018mitigating,gitiaux2021learning}, and
related work accommodates multicategory sensitive attributes
\citep{denis2024fairness}. These interventions can move the target away from
the unconstrained risk minimizer, creating an intrinsic utility--fairness trade-off
\citep{zhao2022inherent,dehdashtian2024ufate}.

A complementary literature quantifies uncertainty in fairness evaluation.
Finite-sample certificates address generalization of learned
representations \citep{gitiaux2021learning}; smoothing and bootstrap methods
support inference for demographic-parity metrics and subgroup audits
\citep{rajkumar2022identification,cherian2024statistical}.
\citet{liu2024empirical} construct empirical-likelihood confidence regions for
a covariance fairness proxy, and \citet{roy2026fairness} study fairness
uncertainty for classifiers trained by constrained stochastic optimization,
using an online multiplier bootstrap.

Sequential fairness research studies delayed impact and runtime monitoring
\citep{liu2018delayed,henzinger2023monitoring}. For independent streams,
stochastic approximation and Polyak--Ruppert averaging support one-pass
estimation at the root-sample-size rate
\citep{robbins1951stochastic,polyak1992acceleration}. Renewable estimating
equations and online covariance methods likewise permit inference without
retaining the full data history
\citep{lin2011aggregated,schifano2016online,chen2020statistical,
han2024online}. Online resampling provides another route to confidence
intervals for stochastic-gradient estimators \citep{fang2018online}, while
\citet{zhu2023online} develop recursive covariance estimation from averaged
stochastic-gradient trajectories.

We study inference when the fairness operator and predictor are learned from
the same stream. The paper makes four contributions.
\begin{enumerate}[(i)]
\item We formulate coefficient inference for fairness-regularized prediction
with an unknown penalty operator. The framework covers logistic and
smoothed-margin classification and linear regression, accounting for the
uncertainty introduced by learning the operator during training.
\item We analyze coupled one-pass updates for the predictor and group--design
covariance. Our expansion reveals how repeated use of each observation in
subsequent penalty gradients creates a first-order fluctuation that survives
Polyak averaging.
\item We derive an augmented central limit theorem retaining score variation,
operator variation, and their signed cross-covariance. Harmonic reuse weights
explain the \mbox{additional} contribution and why treating the learned operator as
fixed can miscalibrate confidence intervals.
\item We construct exact fixed-memory covariance recursions and prove plug-in
consistency and Wald coverage. Four-arm simulations isolate covariance
augmentation, while Adult and IPUMS applications illustrate coefficient
uncertainty and movement along the fairness--utility trade-off.
\end{enumerate}
The theory assumes fixed dimension, independent sampling, and prespecified
designs and penalties; Section \ref{sec:lambda-sequence} discusses vanishing
penalties. Complete proofs and additional results are in the supplement.

\section*{Notation}
For a vector or matrix, \(\norm{\cdot}\) denotes the Euclidean or operator
norm, as appropriate. Write \(a_n\lesssim b_n\) when
\(a_n\leq Cb_n\) for a constant independent of \(n\), and let \(\F_n\) be
the filtration generated by the first \(n\) observations; \(a_n\asymp b_n\)
means both \(a_n\lesssim b_n\) and \(b_n\lesssim a_n\).
The symbols \(\Longrightarrow\) and \(\to_p\) denote convergence in
distribution and probability, respectively. The vectors
\(\boldsymbol1_m\) and matrices \(\boldsymbol I_m\) denote the
dimension-\(m\) all-ones vector and identity matrix, with subscripts omitted
when the dimension is clear; \(\Phi\) is the standard-normal distribution
function. Unless a deterministic sequence is explicitly introduced,
\(\lambda\) is fixed as \(n\to\infty\). The fairness-regularized population
minimizer is \(\boldsymbol\beta_\lambda^*\), with
\(\boldsymbol\beta^*=\boldsymbol\beta_0^*\).

\section{Problem formulation}\label{sec:model}
\renewcommand{\theequation}{2.\arabic{equation}}\setcounter{equation}{0}

Consider an infinite stream of independent copies of
\((\boldsymbol V,Y,Z)\), where \(\boldsymbol V\in\mathcal V\) contains the raw covariates,
\(Y\) is the response, and \(Z\in\{1,\ldots,K\}\), with \(K\ge2\), is the
sensitive attribute. Let
\(\boldsymbol\phi:\mathcal V\to\mathbb R^p\) be a prespecified feature map and
define the design vector
\(\boldsymbol X=\boldsymbol\phi(\boldsymbol V)\). The map may include an
intercept, transformations, interactions, or fixed basis terms. To keep the
online notation compact, write \(\boldsymbol W=(\boldsymbol X,Y,Z)\)
throughout the rest of the paper. Also write
\(Z_g=\mathbbm{1}(Z=g)\),
\(\boldsymbol Z=(Z_1,\ldots,Z_K)^\top\) for the one-hot group vector, and
\(\boldsymbol\pi=\E\boldsymbol Z\), so
\(\pi_g=\Pr(Z=g)\).

\subsection{Threshold classification and a smooth parity surrogate}

The primary problem is binary classification. For
\(\boldsymbol\beta\in\mathbb R^p\), define the decision score and threshold
classifier
\begin{equation}\label{eq:score-classifier}
 \delta_{\boldsymbol\beta}(\boldsymbol V)
 =\boldsymbol\phi(\boldsymbol V)^\top\boldsymbol\beta
 =\boldsymbol X^\top\boldsymbol\beta,
 \qquad
 \widehat Y_{\boldsymbol\beta}
 =\mathbbm{1}\{\delta_{\boldsymbol\beta}(\boldsymbol V)\geq0\}.
\end{equation}
Any fixed nonzero threshold can be absorbed into an intercept. A probability
link may accompany the score; the deployed label follows \eqref{eq:score-classifier}.

We write the observation-level loss and its population risk as
\[
 L(\boldsymbol W,\boldsymbol\beta)
 \quad\text{and}\quad
 \mathcal L(\boldsymbol\beta)=\E\{L(\boldsymbol W,\boldsymbol\beta)\}.
\]
Exact demographic parity requires equal positive-decision probabilities.
Let
\(p_g^+(\boldsymbol\beta)
=\Pr(\widehat Y_{\boldsymbol\beta}=1\mid Z=g)\) for groups with
\(\pi_g>0\). Then
\[
 p_1^+(\boldsymbol\beta)=\cdots=p_K^+(\boldsymbol\beta),
 \qquad
 \Delta_{\rm DP}^{\rm cls}(\boldsymbol\beta)
 =\max_g p_g^+(\boldsymbol\beta)-\min_g p_g^+(\boldsymbol\beta)=0.
\]
Because the threshold indicator makes
\(\Delta_{\rm DP}^{\rm cls}\) nonsmooth and generally nonconvex in
\(\boldsymbol\beta\), we use group mean decision scores as a tractable
surrogate. Define
\[
 \Delta_g^{s}(\boldsymbol\beta)
 =\E\{\delta_{\boldsymbol\beta}(\boldsymbol V)\mid Z=g\}
 -\E\{\delta_{\boldsymbol\beta}(\boldsymbol V)\}.
\]

\begin{condition}\label{Con1}
The feature map \(\boldsymbol\phi\) is fixed before streaming and the decision
score is linear in \(\boldsymbol\beta\), as in
\eqref{eq:score-classifier}. The group coding and loss
\(L(\boldsymbol W,\boldsymbol\beta)\) are also prespecified. Fairness
regularization acts on the decision score, not on the threshold indicator.
\end{condition}

Let \(\boldsymbol\mu_X=\E\boldsymbol X\) and define the population
group--score covariance operator
\begin{equation}\label{eq:C-definition}
 \boldsymbol C
 =\E\left[(\boldsymbol Z-\boldsymbol\pi)
 \{\boldsymbol X-\boldsymbol\mu_X\}^\top\right]
 =\E\left[(\boldsymbol Z-\boldsymbol\pi)
 \boldsymbol X^\top\right]
 \in\mathbb R^{K\times p}.
\end{equation}
For each group,
\[
 \boldsymbol e_g^\top\boldsymbol C\boldsymbol\beta
 =\pi_g\Delta_g^{s}(\boldsymbol\beta).
\]
Here \(\boldsymbol e_g\) is the \(g\)th standard basis vector in
\(\mathbb R^K\).
Consequently,
\[
 \norm{\boldsymbol C\boldsymbol\beta}^2
 =\sum_{g=1}^K\pi_g^2\{\Delta_g^s(\boldsymbol\beta)\}^2,
\]
so the penalty is the sum of prevalence-weighted squared group mean score
deviations.
Thus \(\boldsymbol C\boldsymbol\beta=\boldsymbol0\) if and only if all
positive-probability groups have the same mean decision score. This condition
is the zero set of our smooth convex proxy penalty. Equal mean scores do not
imply equal threshold decision rates, nor does reducing the penalty guarantee
a reduction in the hard-decision DP gap. The theoretical guarantees concern
the score-penalized coefficient target; hard-decision DP is evaluated
separately in the classification applications. Since the rows of
\(\boldsymbol Z-\boldsymbol\pi\) sum to zero, the rows of
\(\boldsymbol C\) are linearly dependent. This redundancy is harmless:
\(\boldsymbol C^\top\boldsymbol C\) and the penalty below are uniquely
defined, and retain a symmetric treatment of the \(K\) groups.

Squaring the surrogate covariances gives the fairness-regularized population
objective
\begin{equation}\label{obj:2}
 \mathcal L^\lambda(\boldsymbol\beta)
 =\mathcal L(\boldsymbol\beta)
 +\frac{\lambda}{2}\norm{\boldsymbol C\boldsymbol\beta}^2,
 \quad
 \boldsymbol\beta_\lambda^*
 =\argmin_{\boldsymbol\beta\in\mathcal B}
 \mathcal L^\lambda(\boldsymbol\beta),
\end{equation}
where \(\lambda\geq0\) determines the utility--fairness trade-off and
\(\mathcal B\subseteq\mathbb R^p\) is a convex parameter space. Under
Condition \ref{Con1}, the penalty term is
\(\lambda\boldsymbol\beta^\top\boldsymbol C^\top\boldsymbol C
\boldsymbol\beta/2\), a convex quadratic function. If the population risk
is strongly convex, adding this term preserves strong convexity. The
inferential target throughout the fixed-penalty theory is
\(\boldsymbol\beta_\lambda^*\), not the unpenalized risk minimizer
\(\boldsymbol\beta^*=\boldsymbol\beta_0^*\).

\subsection{Classification losses}

The threshold rule \eqref{eq:score-classifier} can be fitted with different
convex surrogate losses. The theory depends on the score, loss gradient, and
loss Hessian, rather than on differentiability of the threshold rule.
\begin{enumerate}[(a)]
\item \emph{Logistic regression.} With \(Y\in\{0,1\}\), the fitted class
probability is
\(\{1+\exp(-\boldsymbol X^\top\boldsymbol\beta)\}^{-1}\), and thresholding
that probability at \(1/2\) gives \eqref{eq:score-classifier}. The loss is
\[
 L(\boldsymbol W,\boldsymbol\beta)
 =\log\{1+\exp(\boldsymbol X^\top\boldsymbol\beta)\}
 -Y\boldsymbol X^\top\boldsymbol\beta.
\]
The loss is convex and smooth. The fairness penalty acts on the log-odds
decision score, while exact DP is evaluated from the thresholded
classification decisions.
\item \emph{Smoothed support-vector machines.} Let
\(Y^\pm=2Y-1\in\{-1,1\}\). A twice differentiable convex approximation
\(q_\tau\), with smoothing parameter \(\tau>0\), to the hinge loss is
applied to the signed margin, so
\[
 L(\boldsymbol W,\boldsymbol\beta)
 =q_\tau(Y^\pm\boldsymbol X^\top\boldsymbol\beta)
 +\tfrac\kappa2\norm{\boldsymbol\beta}^2,\qquad \kappa>0.
\]
The reported class label is again \eqref{eq:score-classifier}. The theorem
requires the smooth convex loss and score-level penalty, not a differentiable
threshold map.
\end{enumerate}

A prespecified finite basis is included directly through
\(\boldsymbol\phi\). Methods with learned features or infinite-width neural
tangent kernels introduce additional random or infinite-dimensional objects
and require separate analysis beyond Condition \ref{Con1}.
Equivalent contrast codings of the group mean scores and their common
score-parity zero set are described in the supplement.

\subsection{Linear regression as a special case}

For continuous \(Y\), let the prediction be
\(\widehat Y_{\boldsymbol\beta}
=\boldsymbol\phi(\boldsymbol V)^\top\boldsymbol\beta
=\boldsymbol X^\top\boldsymbol\beta\). In this identity-link case, the same
score condition is no longer a surrogate: \(\boldsymbol C\boldsymbol\beta
=\boldsymbol0\) is exactly equality of group mean predictions. Equivalently,
\[
 \max_g\E(\widehat Y_{\boldsymbol\beta}\mid Z=g)
 -\min_g\E(\widehat Y_{\boldsymbol\beta}\mid Z=g)=0.
\]
We use squared loss,
\[
 L(\boldsymbol W,\boldsymbol\beta)
 =\tfrac12(Y-\boldsymbol X^\top\boldsymbol\beta)^2,
\]
and retain MSE only as the predictive-utility measure. Writing
\(\boldsymbol G=\E(\boldsymbol X\boldsymbol X^\top)\) and
\(\boldsymbol g_Y=\E(\boldsymbol X Y)\), the penalized target is
\begin{equation}\label{eq:linear-regression-target}
 \boldsymbol\beta_\lambda^*
 =\{\boldsymbol G+\lambda
 \boldsymbol C^\top\boldsymbol C\}^{-1}\boldsymbol g_Y,
\end{equation}
when \(\mathcal B=\mathbb R^p\) (or the unconstrained minimizer is interior)
and the matrix is nonsingular.

\section{Methodology}\label{sec3}
\renewcommand{\theequation}{3.\arabic{equation}}\setcounter{equation}{0}

\subsection{Coupled one-pass estimator and oracle benchmark}
\label{sec:coupled-recursion}

The population objective in \eqref{obj:2} is simple when
\(\boldsymbol C\) is known, but the operator must be learned from the same
stream as the predictor. The method therefore has two coupled recursions: one
for the group--design covariance and one for the coefficient.

\medskip\noindent\emph{Population target and streaming moments.}
Fix a penalty value \(\lambda\) for the run and write
\(\boldsymbol M=\boldsymbol C^\top\boldsymbol C\). The estimand is
\(\boldsymbol\beta_\lambda^*\), the minimizer of the population criterion in
\eqref{obj:2}; changing \(\lambda\) changes that estimand. At
time \(n\), estimate the unknown fairness covariance by
\begin{align}\label{eq:online-C-definition}
 \bar{\boldsymbol Z}_n&=n^{-1}\sum_{i=1}^n\boldsymbol Z_i,
 &\bar{\boldsymbol X}_n&=n^{-1}\sum_{i=1}^n
 \boldsymbol X_i,\nonumber\\
 \widehat{\boldsymbol C}_n
 &=n^{-1}\sum_{i=1}^n\boldsymbol Z_i\boldsymbol X_i^\top
 -\bar{\boldsymbol Z}_n\bar{\boldsymbol X}_n^\top,
 &\widehat{\boldsymbol M}_n&=\widehat{\boldsymbol C}_n^\top
 \widehat{\boldsymbol C}_n.
\end{align}
All quantities in \eqref{eq:online-C-definition} are updated by running sums.
Equivalently, the means and cross-moment satisfy
\[
 \bar{\boldsymbol Z}_n=\bar{\boldsymbol Z}_{n-1}
 +n^{-1}(\boldsymbol Z_n-\bar{\boldsymbol Z}_{n-1}),\qquad
 \bar{\boldsymbol X}_n=\bar{\boldsymbol X}_{n-1}
 +n^{-1}(\boldsymbol X_n-\bar{\boldsymbol X}_{n-1}),
\]
with the same recursion for
\(n^{-1}\sum_{i\le n}\boldsymbol Z_i\boldsymbol X_i^\top\). Thus the
operator update stores \(O(Kp)\) numbers and never revisits an observation.

\medskip\noindent\emph{Coefficient update and Polyak average.}
Define \(\boldsymbol\Psi(\boldsymbol\beta,\boldsymbol W)=
\nabla_\beta L(\boldsymbol W,\boldsymbol\beta)\). The update is
\begin{equation}\label{sgd}
 \widehat{\boldsymbol\beta}_n=\widehat{\boldsymbol\beta}_{n-1}
 -\gamma_n\{\boldsymbol\Psi(\widehat{\boldsymbol\beta}_{n-1},
 \boldsymbol W_n)+\lambda\widehat{\boldsymbol M}_n
 \widehat{\boldsymbol\beta}_{n-1}\},
 \quad \gamma_n=\gamma n^{-\alpha},
\end{equation}
where \(\gamma>0\) and \(1/2<\alpha<1\). The reported estimator is the
Polyak average
\(\bar{\boldsymbol\beta}_n=n^{-1}\sum_{i=1}^n
\widehat{\boldsymbol\beta}_i\), updated recursively as
\[
 \bar{\boldsymbol\beta}_n=\bar{\boldsymbol\beta}_{n-1}
 +n^{-1}(\widehat{\boldsymbol\beta}_n-
 \bar{\boldsymbol\beta}_{n-1}).
\]
At iteration \(n\), we observe \(\boldsymbol W_n\), update the
running moments and \(\widehat{\boldsymbol M}_n\), take the stochastic-gradient
step in \eqref{sgd}, and update the average. A fixed compact projection may be
applied after the gradient step to ensure stability. It is asymptotically
inactive when the target lies in the interior.

\medskip\noindent\emph{Oracle comparison and the missing path.}
For comparison only, \(\widehat{\boldsymbol\beta}^{\rm or}_n\) denotes the
infeasible recursion obtained by replacing \(\widehat{\boldsymbol M}_n\) with
the population \(\boldsymbol M\). At the target define
\begin{align}
 \boldsymbol\eta_i
 &=-\{\boldsymbol\Psi(\boldsymbol\beta_\lambda^*,\boldsymbol W_i)
   +\lambda\boldsymbol M\boldsymbol\beta_\lambda^*\},
 \label{eq:eta-def}\\
 \boldsymbol A
&=\nabla^2\mathcal L(\boldsymbol\beta_\lambda^*)+\lambda\boldsymbol M,
 \quad
 \boldsymbol S=\E(\boldsymbol\eta_i\boldsymbol\eta_i^\top).
\end{align}
Thus \(\boldsymbol\eta_i\in\mathbb R^p\) and
\(\boldsymbol A,\boldsymbol S\in\mathbb R^{p\times p}\).
Also put
\(\boldsymbol g(\boldsymbol\beta)=
\E\boldsymbol\Psi(\boldsymbol\beta,\boldsymbol W)
+\lambda\boldsymbol M\boldsymbol\beta\) and
\(\boldsymbol D_n=\widehat{\boldsymbol M}_n-\boldsymbol M\). The population
first-order condition is
\(\boldsymbol g(\boldsymbol\beta_\lambda^*)=\boldsymbol0\). Linearizing the
coefficient error \(\boldsymbol e_n=\widehat{\boldsymbol\beta}_n-
\boldsymbol\beta_\lambda^*\) displays the two stochastic inputs:
\[
 \boldsymbol e_n
 \approx(\boldsymbol I-\gamma_n\boldsymbol A)\boldsymbol e_{n-1}
 +\gamma_n\boldsymbol\eta_n
 -\gamma_n\lambda\boldsymbol D_n\boldsymbol\beta_\lambda^*.
\]
The oracle recursion has only the ordinary score term
\(\boldsymbol\eta_n\). The learned-operator recursion has the additional
term involving \(\boldsymbol D_n\); because observation \(i\) remains in
every later \(\widehat{\boldsymbol M}_n\), this term cannot be treated as
independent one-step noise. Section \ref{sec:covariance-path} shows that its
cumulative contribution is first order after averaging.

\subsection{Regularity conditions and convergence}
\label{sec:regularity}

For \(\boldsymbol\beta\in\mathcal B\), define the centered score
\[
 \boldsymbol\zeta_i(\boldsymbol\beta)
 =\boldsymbol\Psi(\boldsymbol\beta,\boldsymbol W_i)
 -\E\boldsymbol\Psi(\boldsymbol\beta,\boldsymbol W).
\]
\begin{condition}\label{Con2}
For almost every \(\boldsymbol W\), the map
\(\boldsymbol\beta\mapsto L(\boldsymbol W,\boldsymbol\beta)\) is convex. The
penalized mean gradient
\(\boldsymbol g(\boldsymbol\beta)=
\E\boldsymbol\Psi(\boldsymbol\beta,\boldsymbol W)
+\lambda\boldsymbol M\boldsymbol\beta\) has unique zero
\(\boldsymbol\beta_\lambda^*\). On a convex stable set \(\mathcal B\)
containing it, \(\boldsymbol g\) is Lipschitz and strongly monotone.
The dimensions \(p,K\) are fixed. Initialization is deterministic in a fixed
compact convex set \(\mathcal K\subseteq\mathcal B\), with
\(\boldsymbol\beta_\lambda^*\in\operatorname{int}(\mathcal K)\).
The update is projected onto \(\mathcal K\), or its unprojected iterates
remain in \(\mathcal K\) almost surely. These requirements also apply to the
known-operator recursion.
Specifically, for some \(c_0>0\),
\[
 \langle\boldsymbol\beta-\boldsymbol\beta',
 \boldsymbol g(\boldsymbol\beta)-\boldsymbol g(\boldsymbol\beta')\rangle
 \geq c_0\norm{\boldsymbol\beta-\boldsymbol\beta'}^2
 \quad\text{for all }\boldsymbol\beta,\boldsymbol\beta'\in\mathcal B.
\]
\end{condition}

\begin{condition}\label{Con3}
For every \(\boldsymbol\beta\in\mathcal B\),
\(\E\{\boldsymbol\zeta_i(\boldsymbol\beta)\mid\F_{i-1}\}
=\boldsymbol0\) almost surely. For some \(\delta\in(0,2]\), uniformly on
\(\mathcal B\), its conditional \((2+\delta)\)-moment is bounded by
\(C\{1+\norm{\boldsymbol\beta-
\boldsymbol\beta_\lambda^*}^{2+\delta}\}\). The corresponding conditional
Lindeberg condition holds at \(\boldsymbol\beta_\lambda^*\).
\end{condition}

\begin{condition}\label{Con4}
We have \(\min_g\pi_g>0\),
\(\E\norm{\boldsymbol X}^4<\infty\), and
\(\sup_{\lambda\in\Lambda}\norm{\boldsymbol\beta_\lambda^*}<\infty\)
for the compact penalty set \(\Lambda\) under consideration.
\end{condition}

\begin{condition}\label{Con5}
The loss \(L(\boldsymbol W,\boldsymbol\beta)\) is twice continuously
differentiable in \(\boldsymbol\beta\) on \(\mathcal B\).
Put \(q=2+\delta\), with the same \(\delta\) as in Condition \ref{Con3}.
There are random envelopes \(H_0(\boldsymbol W)\) and
\(H_1(\boldsymbol W)\) such that
\[
 \sup_{\boldsymbol\beta\in\mathcal B}
 \norm{\nabla_\beta^2L(\boldsymbol W,\boldsymbol\beta)}
 \leq H_0(\boldsymbol W),\qquad \E H_0(\boldsymbol W)^q<\infty,
\]
and, in a neighborhood of \(\boldsymbol\beta_\lambda^*\),
\[
 \norm{\nabla_\beta^2L(\boldsymbol W,\boldsymbol\beta)
 -\nabla_\beta^2L(\boldsymbol W,\boldsymbol\beta')}
 \leq H_1(\boldsymbol W)\norm{\boldsymbol\beta-\boldsymbol\beta'},
 \qquad \E H_1(\boldsymbol W)<\infty.
\]
Moreover,
\(\E\norm{\boldsymbol\eta_i}^{q}<\infty\),
\(\E\norm{\boldsymbol X}^{2q}<\infty\), and
\(\boldsymbol A\) is positive definite.
\end{condition}

Conditions \ref{Con2}--\ref{Con3} control stochastic approximation;
Condition \ref{Con4} gives the usual sample-covariance moment bounds and
controls the product of the two sample-mean errors; Condition \ref{Con5}
linearizes the loss gradient and supplies the process-level moment bounds.
The endpoint covariance linearization itself uses only the fourth feature
moment in Condition \ref{Con4}. The additional \(4+2\delta\) feature moment
in Condition \ref{Con5} supports the higher-moment recursion and the derived
uniform terminal bound in the functional limit; because any positive
\(\delta\) suffices, an eighth moment is not required. For plug-in
consistency of second moments, Theorem \ref{thm:plugin-consistency} separately
assumes fourth moments of \(\boldsymbol\eta_i\) and
\(\boldsymbol\rho_i\).

These conditions concern the loss and fixed design representation.
For logistic regression and other smooth generalized-linear-model losses,
they reduce to moments of \(\boldsymbol X\), integrability of the score and
Hessian envelopes, and nonsingularity of the penalized information matrix.
Linear regression additionally needs the corresponding response--design
moments; its Hessian is constant. A smoothed hinge has a bounded response
contribution and a controlled smoothness parameter. An adaptively learned
design representation or a dimension increasing with \(n\) would add another
estimated nuisance and lies outside our scope.

\begin{remark}\label{rem:stability}
Under Condition \ref{Con2}, compact projection ensures bounded iterates while
the target is interior. The supplement proves almost-sure convergence and
eventual projection inactivity before applying the linearized recursion.
For logistic and linear regression, the conditions can be checked from the
model-specific score, Hessian, feature moments, positive group probabilities,
and positive definiteness of the population information matrix. For the
smoothed SVM, the additional \(L_2\) term supplies strong convexity when the
smooth margin loss alone does not.
\end{remark}

\begin{theorem}[Iterate and average rates]\label{order}
Under Conditions \ref{Con1}--\ref{Con4}, for \(m=1,2\),
\[
 \E\norm{\widehat{\boldsymbol\beta}_n-
 \boldsymbol\beta_\lambda^*}^m=O(\gamma_n^{m/2}).
\]
If Condition \ref{Con5} also holds, the same iterate bound holds for
\(m=q=2+\delta\), and
\[
 \norm{\bar{\boldsymbol\beta}_n-
 \boldsymbol\beta_\lambda^*}=O_p(n^{-1/2}).
\]
\end{theorem}

The two rates describe different parts of the online procedure. The last
iterate is updated with step size \(\gamma_n\), so its stochastic
neighborhood has radius \(\gamma_n^{1/2}\). Polyak averaging converts the
accumulated score noise to the root-\(n\) scale. Although the learned operator
is pointwise root-\(n\) consistent, observation \(i\) remains in every later
operator estimate and affects many later penalty gradients. Its cumulative
effect on the average can therefore be first order even when its effect on one
update is small.

Strong monotonicity contracts the
previous error, the stochastic loss score supplies the usual
\(O(\gamma_n^2)\) forcing, and operator estimation contributes
\(O(\gamma_n n^{-1})\). These terms give the last-iterate rate without a
fourth-power recursion. Summation by parts then turns the averaged recursion
into an empirical score sum, a nested operator sum, and boundary remainders.
Condition \ref{Con4} controls the second-order operator remainder; Condition
\ref{Con5} controls the gradient linearization and Lindeberg step. The
supplement gives each bound and shows that eighth primitive moments are
unnecessary.

If the operator were known, ordinary Polyak inference would be valid:
\begin{proposition}[Known-operator benchmark]\label{prop:known-operator}
Under Conditions \ref{Con1}--\ref{Con5},
\[
 \sqrt n(\bar{\boldsymbol\beta}^{\rm or}_n-
 \boldsymbol\beta_\lambda^*)\ \Longrightarrow\
 N(\boldsymbol0,\boldsymbol A^{-1}\boldsymbol S\boldsymbol A^{-1}).
\]
\end{proposition}

\subsection{The covariance-estimation path and endpoint uncertainty}
\label{sec:covariance-path}

Let \(\boldsymbol\pi=\E\boldsymbol Z\) and
\(\boldsymbol\mu_X=\E\{\boldsymbol X\}\). The first-order influence
of observation \(i\) on the fairness covariance is
\begin{equation}\label{eq:Qrho-def}
 \boldsymbol Q_i=(\boldsymbol Z_i-\boldsymbol\pi)
 \{\boldsymbol X_i-\boldsymbol\mu_X\}^{\top}-\boldsymbol C,
 \quad
 \boldsymbol\rho_i=
 (\boldsymbol C^\top\boldsymbol Q_i+
 \boldsymbol Q_i^\top\boldsymbol C)\boldsymbol\beta_\lambda^*.
\end{equation}
Here \(\boldsymbol Q_i\in\mathbb R^{K\times p}\) and
\(\boldsymbol\rho_i\in\mathbb R^p\).
The sample covariance has the expansion
\begin{equation}\label{lem:C-linear}
 \widehat{\boldsymbol C}_i-\boldsymbol C
 =i^{-1}\sum_{k=1}^i\boldsymbol Q_k+
 \boldsymbol R_{C,i},
 \quad
 \boldsymbol R_{C,i}=-(\bar{\boldsymbol Z}_i-\boldsymbol\pi)
 (\bar{\boldsymbol X}_i-\boldsymbol\mu_X)^\top .
\end{equation}
Under Condition \ref{Con4},
\(\E\norm{\boldsymbol R_{C,i}}^2=O(i^{-2})\), whereas
\(\E\norm{\widehat{\boldsymbol C}_i-\boldsymbol C}^2=O(i^{-1})\).
Writing \(\boldsymbol\Delta_i=\widehat{\boldsymbol C}_i-\boldsymbol C\)
and using
\[
 \widehat{\boldsymbol M}_i-\boldsymbol M
 =\boldsymbol C^\top\boldsymbol\Delta_i+
 \boldsymbol\Delta_i^\top\boldsymbol C+
 \boldsymbol\Delta_i^\top\boldsymbol\Delta_i
\]
therefore gives
\begin{equation}\label{lem:M-linear}
 (\widehat{\boldsymbol M}_i-\boldsymbol M)
 \boldsymbol\beta_\lambda^*
 =i^{-1}\sum_{k=1}^i\boldsymbol\rho_k+
 \boldsymbol r_{M,i},
 \quad \E\norm{\boldsymbol r_{M,i}}=O(i^{-1}).
\end{equation}
Consequently,
\(n^{-1/2}\sum_{i=1}^n\norm{\boldsymbol r_{M,i}}=o_p(1)\).
The supplement proves the displayed moment bounds from the fourth feature
moment and writes \(\boldsymbol r_{M,i}\) explicitly. The product of the two
sample-mean errors and the quadratic term
\(\boldsymbol\Delta_i^\top\boldsymbol\Delta_i\) are second order, while the
linear influence \(\boldsymbol\rho_i\) survives after averaging.

Put \(\boldsymbol D_i=\widehat{\boldsymbol M}_i-\boldsymbol M\) and define
the stable product
\[
 \boldsymbol P_j^i=\prod_{\ell=j+1}^i
 (\boldsymbol I-\gamma_\ell\boldsymbol A),
 \quad \boldsymbol P_i^i=\boldsymbol I,
\]
together with the infeasible covariance-estimation path
\begin{equation}\label{eq:bias-path}
 \boldsymbol{\mathfrak b}_i=-\lambda\sum_{j=1}^i
 \boldsymbol P_j^i\gamma_j\boldsymbol D_j
 \boldsymbol\beta_\lambda^*.
\end{equation}

\begin{theorem}[Infeasible functional central limit theorem]\label{fclt}
Under Conditions \ref{Con1}--\ref{Con5},
\begin{equation}\label{eq:infeasible-fclt}
 \boldsymbol\Phi_n^{\rm adj}(r)
 =\frac1{\sqrt n}\sum_{i=1}^{\floor{nr}}
 \{\widehat{\boldsymbol\beta}_i-
 \boldsymbol\beta_\lambda^*-\boldsymbol{\mathfrak b}_i\}
 \ \Longrightarrow\
 \boldsymbol A^{-1}\boldsymbol S^{1/2}\boldsymbol W_p(r),
 \quad 0\le r\le1,
\end{equation}
where \(\boldsymbol W_p\) is a standard \(p\)-dimensional Brownian motion.
\end{theorem}

Theorem \ref{fclt} has the same Brownian limit as the known-operator
recursion only after subtracting \(\boldsymbol{\mathfrak b}_i\). The average
path displacement
\[
 \operatorname{path}_n(\boldsymbol\beta_\lambda^*)
 =-n^{-1}\sum_{i=1}^n\boldsymbol{\mathfrak b}_i
 =\frac{\lambda}{n}\sum_{i=1}^n\sum_{j=1}^i
 \boldsymbol P_j^i\gamma_j\boldsymbol D_j
 \boldsymbol\beta_\lambda^*
\]
is \(O_p(\lambda n^{-1/2})\). For fixed \(\lambda>0\), it is therefore
of the same order as the ordinary score fluctuation, even though
\(\widehat{\boldsymbol C}_i\) is itself a consistent covariance estimator.
The reason is repeated use: observation \(j\) enters every later
\(\widehat{\boldsymbol M}_i\), so its small one-step effects accumulate along
the entire stochastic-gradient path.

The path is infeasible because it contains \(\boldsymbol A\),
\(\boldsymbol M\), and \(\boldsymbol\beta_\lambda^*\). The supplement proves
the process result from the stated conditions, including the maximal terminal
remainder bound needed for tightness; that bound is derived from the
\((2+\delta)\)-moment recursion and is not imposed as an additional
assumption. The feasible procedure below keeps the raw Polyak average as its
center and transfers the path contribution to the endpoint covariance.

For endpoint inference, reverse the nested sums. Let
\(h_{i,n}=\sum_{j=i}^n j^{-1}\). Their empirical second moment is bounded,
although the earliest weights grow logarithmically; this records the repeated
use of early observations.

\begin{theorem}[Joint endpoint expansion]\label{thm:joint-expansion}
Under Conditions \ref{Con1}--\ref{Con5},
\begin{equation}\label{eq:joint-expansion}
 \sqrt n(\bar{\boldsymbol\beta}_n-
 \boldsymbol\beta_\lambda^*)
 =\boldsymbol A^{-1}\frac1{\sqrt n}\sum_{i=1}^n
 \{\boldsymbol\eta_i-\lambda h_{i,n}
 \boldsymbol\rho_i\}+o_p(1).
\end{equation}
\end{theorem}

Write
\(\boldsymbol\Gamma_{\eta\rho}=
\E(\boldsymbol\eta_i\boldsymbol\rho_i^\top)\) and
\(\boldsymbol\Gamma_{\rho\rho}=
\E(\boldsymbol\rho_i\boldsymbol\rho_i^\top)\). Since
\(n^{-1}\sum_i h_{i,n}\to1\) and
\(n^{-1}\sum_i h_{i,n}^2\to2\), the endpoint covariance is
\begin{equation}\label{eq:omega-aug}
 \boldsymbol\Omega_{\rm aug}
 =\boldsymbol S-\lambda\{\boldsymbol\Gamma_{\eta\rho}+
 \boldsymbol\Gamma_{\eta\rho}^\top\}
 +2\lambda^2\boldsymbol\Gamma_{\rho\rho},
 \quad
 \boldsymbol\Sigma_{\rm aug}=
 \boldsymbol A^{-1}\boldsymbol\Omega_{\rm aug}\boldsymbol A^{-1}.
\end{equation}

\begin{theorem}[Augmented central limit theorem]\label{thm:augmented-clt}
Under Conditions \ref{Con1}--\ref{Con5},
\[
 \sqrt n(\bar{\boldsymbol\beta}_n-
 \boldsymbol\beta_\lambda^*)\Longrightarrow
 N(\boldsymbol0,\boldsymbol\Sigma_{\rm aug}).
\]
\end{theorem}

The ordinary and augmented covariance matrices can be compared directly:
\begin{align}
 \boldsymbol\Sigma_{\rm ord}
 &=\boldsymbol A^{-1}\boldsymbol S\boldsymbol A^{-1},\nonumber\\
 \boldsymbol\Sigma_{\rm aug}
 &=\boldsymbol\Sigma_{\rm ord}
 -\lambda\boldsymbol A^{-1}
   (\boldsymbol\Gamma_{\eta\rho}+
    \boldsymbol\Gamma_{\eta\rho}^\top)\boldsymbol A^{-1}
 +2\lambda^2\boldsymbol A^{-1}
   \boldsymbol\Gamma_{\rho\rho}\boldsymbol A^{-1}.
 \label{eq:ordinary-versus-augmented}
\end{align}
The ordinary formula is the correct benchmark when the fairness operator is
known; using it with a learned operator sets both additional terms in
\eqref{eq:ordinary-versus-augmented} to zero. The signed cross term records
that the loss score and fairness-operator influence are computed from the same
observation. The final term is positive semidefinite and records repeated use
of that observation in all later operator estimates.

This comparison also explains the undercoverage seen at large penalties. For
coordinate \(j\), let \(v_{{\rm ord},j}\) and \(v_{{\rm aug},j}\) be the two
diagonal variances. If a nominal 95\% interval uses the ordinary standard
error while the augmented law is true, its Gaussian coverage is approximately
\[
 2\Phi\!\left(1.96\sqrt{v_{{\rm ord},j}/v_{{\rm aug},j}}\right)-1.
\]
It is below 0.95 whenever the ordinary variance is too small. As \(\lambda\)
increases, the omitted operator contribution contains the nonnegative
quadratic term \(2\lambda^2\boldsymbol\Gamma_{\rho\rho}\); unless it is
degenerate or offset by the signed cross term and the \(\lambda\)-dependent
curvature \(\boldsymbol A^{-1}\), it eventually dominates and ordinary
intervals become too narrow. Thus undercoverage is the typical large-penalty
pattern, although the variance difference need not be monotone for every
coordinate or design.

The coefficient 2 in \eqref{eq:omega-aug} is created by repeated online use
of the same observation. Substitution of the operator expansion into the
Polyak reduction gives the exact reversal
\[
 \sum_{i=1}^n i^{-1}\sum_{k=1}^i\boldsymbol\rho_k
 =\sum_{k=1}^n\left(\sum_{i=k}^ni^{-1}\right)\boldsymbol\rho_k
 =\sum_{k=1}^nh_{k,n}\boldsymbol\rho_k.
\]
The identities \(n^{-1}\sum_kh_{k,n}=1\) and
\(n^{-1}\sum_kh_{k,n}^2\to2\) generate the cross and quadratic terms in
the augmented covariance. Because \(\boldsymbol\eta_k\) and
\(\boldsymbol\rho_k\) use the same observation, their cross covariance cannot
be discarded. Thus the augmentation is not simply a positive variance
inflation; the signed cross term is part of the first-order law.

When \(\lambda=0\), both operator
terms vanish and ordinary Polyak inference is recovered. When the population
operator is supplied, \(\boldsymbol\rho_i\) is absent and Proposition
\ref{prop:known-operator} applies. When the operator is learned and
\(\lambda>0\), ordinary and augmented procedures use the same coefficient
center but generally have different first-order covariance. The four-arm
experiment tests each of these implications.

\subsection{Choice of the fairness-penalty sequence}
\label{sec:lambda-sequence}

The preceding results treat \(\lambda\) as fixed and target
\(\boldsymbol\beta_\lambda^*\), as appropriate when the utility--fairness
trade-off defines the population rule. If inference instead targets the
unpenalized \(\boldsymbol\beta^*=\boldsymbol\beta_0^*\), put
\(\boldsymbol H^*=\nabla^2\mathcal L(\boldsymbol\beta^*)\) and
\(\boldsymbol S_0=\operatorname{Var}\{
\boldsymbol\Psi(\boldsymbol\beta^*,\boldsymbol W)\}\). If
\(\boldsymbol\beta^*\) is interior and \(\boldsymbol H^*\) is nonsingular,
the implicit-function theorem gives
\begin{equation}\label{eq:lambda-target-expansion}
 \boldsymbol\beta_\lambda^*-\boldsymbol\beta^*
 =-\lambda(\boldsymbol H^*)^{-1}\boldsymbol M
 \boldsymbol\beta^*+o(\lambda).
\end{equation}

\begin{corollary}[Fairness-penalty regimes]\label{cor:lambda-regimes}
Under Conditions \ref{Con1}--\ref{Con5}, the fixed-\(\lambda\) conclusion of
Theorem \ref{thm:augmented-clt} holds. Suppose additionally that
\eqref{eq:lambda-target-expansion} and the joint endpoint expansion hold
uniformly along a deterministic sequence \(\lambda_n\to0\), and that the
corresponding curvature and augmented influence covariance converge to
\(\boldsymbol H^*\) and \(\boldsymbol S_0\), respectively. Then:
\begin{enumerate}[(i)]
\item for fixed \(\lambda>0\), the Polyak average is root-\(n\) consistent
and asymptotically normal for \(\boldsymbol\beta_\lambda^*\), with augmented
covariance \(\boldsymbol\Sigma_{\rm aug}\);
\item if \(\lambda_n=o(n^{-1/2})\), then
\[
 \sqrt n(\bar{\boldsymbol\beta}_n-\boldsymbol\beta^*)
 \Longrightarrow
 N\{\boldsymbol0,(\boldsymbol H^*)^{-1}
 \boldsymbol S_0(\boldsymbol H^*)^{-1}\};
\]
\item if \(\sqrt n\lambda_n\to c\in(0,\infty)\), the covariance in part
(ii) is unchanged, but the limiting mean is
\[
 -c(\boldsymbol H^*)^{-1}\boldsymbol M\boldsymbol\beta^*;
\]
\item if \(\sqrt n\lambda_n\to\infty\) while \(\lambda_n\to0\), the Polyak
average remains consistent for \(\boldsymbol\beta^*\), but no centered
root-\(n\) limit exists unless
\(\boldsymbol M\boldsymbol\beta^*=\boldsymbol0\).
\end{enumerate}
\end{corollary}

Thus a fixed positive penalty defines a fairness-regularized population
target. Root-\(n\) inference for the unpenalized target requires a vanishing
penalty; the supplement proves the corollary and records where uniformity
along the penalty sequence is used.

\section{Online augmented plug-in inference}\label{sec:inference}
\renewcommand{\theequation}{4.\arabic{equation}}\setcounter{equation}{0}

The endpoint covariance can be estimated without storing observations. For
\(i\ge2\), all influence contributions use nuisance estimates available
immediately before observation \(i\) arrives:
\begin{align}\label{eq:feasible-influences}
 \widehat{\boldsymbol\eta}_i
 &=-\{\boldsymbol\Psi(\widehat{\boldsymbol\beta}_{i-1},
 \boldsymbol W_i)+\lambda\widehat{\boldsymbol M}_{i-1}
 \widehat{\boldsymbol\beta}_{i-1}\},\\
 \widehat{\boldsymbol Q}_i
 &=(\boldsymbol Z_i-\bar{\boldsymbol Z}_{i-1})
 \{\boldsymbol X_i-\bar{\boldsymbol X}_{i-1}\}^{\top}
 -\widehat{\boldsymbol C}_{i-1},\nonumber\\
 \widehat{\boldsymbol\rho}_i
 &=\{\widehat{\boldsymbol C}_{i-1}^\top\widehat{\boldsymbol Q}_i+
 \widehat{\boldsymbol Q}_i^\top\widehat{\boldsymbol C}_{i-1}\}
 \widehat{\boldsymbol\beta}_{i-1}.\nonumber
\end{align}
Lagging makes the nuisance state predictable and prevents an observation from
entering its own influence estimate. The optimization update \eqref{sgd}
still uses the current operator. At \(i=1\), the three influence
contributions are initialized at zero. Define
\[
 \boldsymbol d_{Z,i}=\boldsymbol Z_i-\bar{\boldsymbol Z}_{i-1},
 \quad
 \boldsymbol d_{X,i}=\boldsymbol X_i-
 \bar{\boldsymbol X}_{i-1}.
\]
The sample covariance in \eqref{eq:online-C-definition} has the exact
Welford-type recursion
\begin{align}\label{eq:online-C-updates}
 \bar{\boldsymbol Z}_i
 &=\bar{\boldsymbol Z}_{i-1}+i^{-1}\boldsymbol d_{Z,i},
 &\bar{\boldsymbol X}_i
 &=\bar{\boldsymbol X}_{i-1}+i^{-1}\boldsymbol d_{X,i},\\
 \widehat{\boldsymbol C}_i
 &=\frac{i-1}{i}\widehat{\boldsymbol C}_{i-1}
 +\frac{i-1}{i^2}\boldsymbol d_{Z,i}
 \boldsymbol d_{X,i}^{\top},
 &\widehat{\boldsymbol M}_i
 &=\widehat{\boldsymbol C}_i^\top\widehat{\boldsymbol C}_i.\nonumber
\end{align}
Thus the fairness state requires only \(O(Kp)\) storage.
The influence estimates in \eqref{eq:feasible-influences} use the preceding state,
whereas \eqref{sgd} uses the current \(\widehat{\boldsymbol M}_i\).

The Polyak average and Hessian estimate obey
\begin{align}\label{eq:online-beta-A}
 \bar{\boldsymbol\beta}_i
 &=\bar{\boldsymbol\beta}_{i-1}
 +i^{-1}(\widehat{\boldsymbol\beta}_i-
 \bar{\boldsymbol\beta}_{i-1}),\\
 \boldsymbol{\mathcal A}_i
 &=\boldsymbol{\mathcal A}_{i-1}+
 \nabla_\beta^2L(\boldsymbol W_i,\widehat{\boldsymbol\beta}_{i-1})
 +\lambda\widehat{\boldsymbol M}_i,
 \quad \widehat{\boldsymbol A}_i=i^{-1}\boldsymbol{\mathcal A}_i.\nonumber
\end{align}

Define the harmonic weight \(h_{j,i}=\sum_{k=j}^ik^{-1}\). To update the
harmonic covariance without storing past observations, set
\begin{align*}
 \boldsymbol{\mathcal G}_i
 &=\sum_{j=1}^i\widehat{\boldsymbol\eta}_j
   \widehat{\boldsymbol\eta}_j^\top,
 &\boldsymbol{\mathcal E}^{(0)}_i
 &=\sum_{j=1}^i\widehat{\boldsymbol\eta}_j
   \widehat{\boldsymbol\rho}_j^\top,\\
 \boldsymbol{\mathcal E}^{(1)}_i
 &=\sum_{j=1}^ih_{j,i}\widehat{\boldsymbol\eta}_j
   \widehat{\boldsymbol\rho}_j^\top,
 &\boldsymbol{\mathcal R}^{(q)}_i
 &=\sum_{j=1}^ih_{j,i}^q\widehat{\boldsymbol\rho}_j
   \widehat{\boldsymbol\rho}_j^\top,\quad q=0,1,2.
\end{align*}
Because \(h_{j,i}=h_{j,i-1}+i^{-1}\) for \(j<i\), with
\(h_{i,i}=i^{-1}\), these quantities satisfy
\begin{align*}
 \boldsymbol{\mathcal G}_i
 &=\boldsymbol{\mathcal G}_{i-1}
   +\widehat{\boldsymbol\eta}_i\widehat{\boldsymbol\eta}_i^\top,\\
 \boldsymbol{\mathcal E}^{(0)}_i
 &=\boldsymbol{\mathcal E}^{(0)}_{i-1}
   +\widehat{\boldsymbol\eta}_i\widehat{\boldsymbol\rho}_i^\top,
 &\boldsymbol{\mathcal E}^{(1)}_i
 &=\boldsymbol{\mathcal E}^{(1)}_{i-1}
   +i^{-1}\boldsymbol{\mathcal E}^{(0)}_i,\\
 \boldsymbol{\mathcal R}^{(0)}_i
 &=\boldsymbol{\mathcal R}^{(0)}_{i-1}
   +\widehat{\boldsymbol\rho}_i\widehat{\boldsymbol\rho}_i^\top,
 &\boldsymbol{\mathcal R}^{(1)}_i
 &=\boldsymbol{\mathcal R}^{(1)}_{i-1}
   +i^{-1}\boldsymbol{\mathcal R}^{(0)}_i,\\
 \boldsymbol{\mathcal R}^{(2)}_i
 &=\boldsymbol{\mathcal R}^{(2)}_{i-1}
   +2i^{-1}\boldsymbol{\mathcal R}^{(1)}_{i-1}
   +i^{-2}\boldsymbol{\mathcal R}^{(0)}_i.
\end{align*}
All states start at zero; using the updated lower-order state in each line
incorporates the new observation and the increment to every existing harmonic
weight. For example, the three weighted updates follow by expanding the new
harmonic weight before collecting the current observation:
\begin{align*}
 \boldsymbol{\mathcal E}^{(1)}_i
 &=\sum_{j<i}(h_{j,i-1}+i^{-1})
   \widehat{\boldsymbol\eta}_j\widehat{\boldsymbol\rho}_j^\top
   +i^{-1}\widehat{\boldsymbol\eta}_i
   \widehat{\boldsymbol\rho}_i^\top
  =\boldsymbol{\mathcal E}^{(1)}_{i-1}
   +i^{-1}\boldsymbol{\mathcal E}^{(0)}_i,\\
 \boldsymbol{\mathcal R}^{(1)}_i
 &=\boldsymbol{\mathcal R}^{(1)}_{i-1}
   +i^{-1}\boldsymbol{\mathcal R}^{(0)}_i,\\
 \boldsymbol{\mathcal R}^{(2)}_i
 &=\sum_{j<i}(h_{j,i-1}+i^{-1})^2
   \widehat{\boldsymbol\rho}_j\widehat{\boldsymbol\rho}_j^\top
   +i^{-2}\widehat{\boldsymbol\rho}_i
   \widehat{\boldsymbol\rho}_i^\top\\
 &=\boldsymbol{\mathcal R}^{(2)}_{i-1}
   +2i^{-1}\boldsymbol{\mathcal R}^{(1)}_{i-1}
   +i^{-2}\boldsymbol{\mathcal R}^{(0)}_i.
\end{align*}
Thus no past influence vector or harmonic weight is retained. Define the
effective endpoint contribution
\[
 \widehat{\boldsymbol\xi}_{j,i}
 =\widehat{\boldsymbol\eta}_j
  -\lambda h_{j,i}\widehat{\boldsymbol\rho}_j.
\]
Then \(i^{-1}\sum_{j\le i}\widehat{\boldsymbol\xi}_{j,i}
\widehat{\boldsymbol\xi}_{j,i}^\top\) expands into the three accumulated
terms below. The endpoint estimator is therefore
\begin{align}\label{eq:omega-harmonic}
 \widehat{\boldsymbol\Omega}^{\rm harm}_i
 &=i^{-1}\left[\boldsymbol{\mathcal G}_i
 -\lambda\{\boldsymbol{\mathcal E}^{(1)}_i+
 \boldsymbol{\mathcal E}^{(1)\top}_i\}
 +\lambda^2\boldsymbol{\mathcal R}^{(2)}_i\right],\\
 \widehat{\boldsymbol\Sigma}^{\rm aug}_i
 &=\widehat{\boldsymbol A}_i^{-1}
\widehat{\boldsymbol\Omega}^{\rm harm}_i
 \widehat{\boldsymbol A}_i^{-\top}.\nonumber
\end{align}
The inverse is formed whenever \(\widehat{\boldsymbol A}_i\) is nonsingular;
Theorem \ref{thm:plugin-consistency} and positive definiteness of
\(\boldsymbol A\) ensure this event has probability tending to one.
The active state consists of one \(K\times p\) operator and a fixed number
of \(p\times p\) matrices, so memory is \(O(Kp+p^2)\), independent of stream
length. Each observation requires the loss gradient and Hessian plus these
matrix updates; inversion is needed only when inference is reported. The
supplement verifies the recursion algebra and gives the complete one-pass
pseudocode as Algorithm S1.

\begin{theorem}[Plug-in consistency]\label{thm:plugin-consistency}
Suppose Conditions \ref{Con1}--\ref{Con5} hold and
\(\E(\norm{\boldsymbol\eta_i}^4+\norm{\boldsymbol\rho_i}^4)<\infty\).
Then \(\widehat{\boldsymbol A}_n\to_p\boldsymbol A\),
\(\widehat{\boldsymbol\Omega}^{\rm harm}_n\to_p
\boldsymbol\Omega_{\rm aug}\), and
\(\widehat{\boldsymbol\Sigma}^{\rm aug}_n\to_p
\boldsymbol\Sigma_{\rm aug}\).
\end{theorem}

For \(0<a<1\), let \(z_{1-a/2}\) be the corresponding standard-normal
quantile. For coordinate \(j\), the proposed \((1-a)\) interval is
\begin{equation}\label{CIplugin-aug}
 \bar\beta_{n,j}\ \mathbin{\pm}\ z_{1-a/2}
 \left\{(\widehat{\boldsymbol\Sigma}^{\rm aug}_n)_{jj}/n\right\}^{1/2}.
\end{equation}
\begin{corollary}[Wald coverage]\label{cor:plugin}
Under Theorems \ref{thm:augmented-clt} and
\ref{thm:plugin-consistency}, if
\((\boldsymbol\Sigma_{\rm aug})_{jj}>0\), interval
\eqref{CIplugin-aug} has limiting coverage \(1-a\) for
\(\beta_{\lambda,j}^*\).
\end{corollary}

\section{Simulation studies}\label{sec4}
\renewcommand{\theequation}{5.\arabic{equation}}\setcounter{equation}{0}

The simulation is organized around three implications of the theory:
convergence to the penalty-specific target, root-\(n\) interval contraction,
and the additional uncertainty created by estimating the fairness operator.
Results are recorded every 10,000 observations through 300,000; the main text
reports checkpoints at 50,000, 150,000, and 300,000, with complete checkpoint
tables and further trajectories in the supplement.

\subsection{Design and identifying arms}

For \(K\in\{2,5\}\) and \(d\in\{10,40\}\), let
\(\boldsymbol v_d=d^{-1/2}\boldsymbol1_d\) and assign equally spaced group
scores \(a_g\in[-1,1]\), with
\(\bar a=\sum_g\pi_ga_g\). Conditional on group \(g\),
\[
 \boldsymbol X=1.25(a_g-\bar a)\boldsymbol v_d
 +U\boldsymbol v_d+\boldsymbol\varepsilon,
 \quad U\sim N(0,1),\quad
 \boldsymbol\varepsilon\sim N(\boldsymbol0,\boldsymbol I_d).
\]
The latent variable \(U\) and noise vector \(\boldsymbol\varepsilon\) are
mutually independent and independent of group membership.
An intercept is appended for fitting, so \(d\) counts non-intercept features
and the coefficient dimension is \(p=d+1\).
Binary outcomes follow a logistic law in
\(\boldsymbol v_d^\top\boldsymbol X\) and are fitted with logistic-regression
and smooth-hinge losses. The continuous outcome is fitted by linear regression
and additionally contains a group-by-\(X_2\) interaction. In every model the
fairness penalty acts on the decision score
\(\boldsymbol X^\top\boldsymbol\beta\). Logistic regression maps this score
through the logistic link, the smoothed SVM uses a smooth margin loss, and
linear regression uses the identity map. Consequently, the regression
fairness metric is the maximum minus the minimum group mean prediction, never
a subgroup MSE contrast. The supplement details the outcome laws and targets.

Every setting uses the raw-covariance operator, a fixed numerical ridge of
0.2 absorbed into the fitted loss, projection radius 2, and
\(\lambda\in\{0,0.1,0.5,1,4\}\). The step sequence is
\(0.8(i+1000)^{-0.60}\), and reported Polyak averages discard the first 5,000
observations. Each of the 60 model--group--dimension--penalty configurations
has 200 independent streams of length 300,000. All 7,200 setting--arm--
checkpoint summaries have 200 successful replications and finite outputs.

Let \(\bar{\boldsymbol\beta}_n^{\rm K}\) denote the known-operator average,
\(\bar{\boldsymbol\beta}_n\) the learned-operator average, and
\(\bar{\boldsymbol{\mathfrak b}}_n=n^{-1}\sum_{i=1}^n
\boldsymbol{\mathfrak b}_i\). The four arms in Table \ref{tab:four-arms}
separate center and covariance effects.

\begin{table}[!htbp]
\centering\small
\caption{Four-arm design. Arms 2 and 3 have exactly the same learned-operator
center, so their difference isolates covariance augmentation.}
\label{tab:four-arms}
\begin{tabular}{@{}c>{\centering\arraybackslash}p{.19\textwidth}>{\centering\arraybackslash}p{.22\textwidth}>{\raggedright\arraybackslash}p{.45\textwidth}@{}}
\toprule
Arm & Reported center & Reference covariance & Statistical role\\
\midrule
1 & \(\bar{\boldsymbol\beta}_n^{\rm K}\)
& \(\boldsymbol\Sigma_{\rm ord}\) & Known-operator benchmark\\
2 & \(\bar{\boldsymbol\beta}_n\)
& \(\boldsymbol\Sigma_{\rm ord}\) & Learned operator; path term omitted\\
3 & \(\bar{\boldsymbol\beta}_n\)
& \(\boldsymbol\Sigma_{\rm aug}\) & Proposed feasible procedure\\
4 & \(\bar{\boldsymbol\beta}_n-\bar{\boldsymbol{\mathfrak b}}_n\)
& \(\boldsymbol\Sigma_{\rm ord}\) & Infeasible path-adjusted benchmark\\
\bottomrule
\end{tabular}
\end{table}
\FloatBarrier

\subsection{Convergence to the penalty-specific target}

\begin{figure}[!ht]
\centering
\includegraphics[width=.58\textwidth]{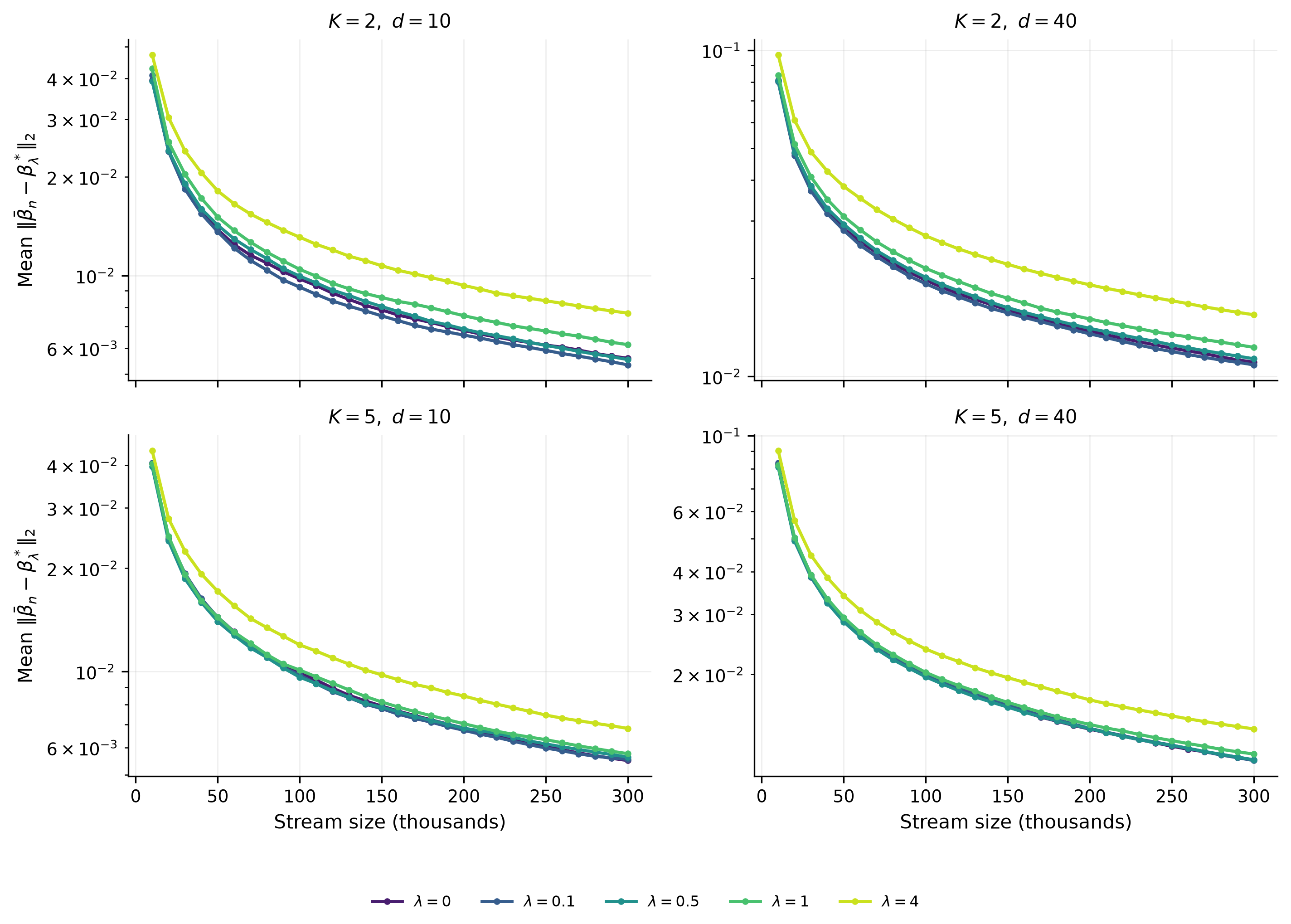}
\caption{Logistic-regression Arm-3 mean distance to the penalty-specific
target across 200 streams at all 30 checkpoints, for every $\lambda$, group
count, and dimension.}
\label{fig:distance-logistic}
\end{figure}

In Figure \ref{fig:distance-logistic}, each curve is the mean Euclidean
distance from the learned-operator Polyak average to its population penalized
minimizer. Using an unpenalized target would conflate optimization error with
penalty-induced estimand movement.

\FloatBarrier

Distances decrease in all four panels. The \(d=40\) curves start farther from
their targets because more coordinates are estimated, while comparable slopes
across \(K\), \(d\), and \(\lambda\) show that learning the fairness operator
does not prevent target tracking. The supplement reports numerical means,
between-replication SDs, and the corresponding smoothed-SVM and
linear-regression trajectories.

\subsection{Coordinate convergence, precision, and coverage}

Across the full grid, every Arm-3 mean Euclidean distance decreases at each
reported checkpoint. The tables below assess calibrated coordinatewise
uncertainty; representative coordinate paths and penalty-specific oracle
targets are retained in the supplement.

Tables \ref{tab:simulation-asymptotic}--
\ref{tab:simulation-asymptotic-squared} report Monte Carlo mean coverage and
interval length separately for each model; between-stream SDs are in the
supplement. Across losses, group counts, and penalties,
augmented coverage stays close to the nominal level from the early checkpoint
onward, while interval lengths contract regularly with stream size. Their
contraction closely follows the root-\(n\) scale predicted by the
central limit theorem, and calibration does not deteriorate at the larger
penalties where operator uncertainty matters most. The similar behavior of
the logistic, smooth-margin, and squared losses supports the loss-generic
conditions rather than a model-specific covariance calculation. The main
paper reports \(d=10\); the supplement contains all higher-dimensional
results.

\begin{table}[!htbp]
\spacingset{1.0}
\centering
\caption{Arm-3 augmented 95\% interval calibration, $d=10$: Logistic regression loss. Entries are Monte Carlo mean coordinatewise coverage and mean interval length ($\times100$) across 200 streams.}
\label{tab:simulation-asymptotic}
\small
\renewcommand{\arraystretch}{1.00}
\setlength{\tabcolsep}{3pt}
\begin{tabular}{@{}rrrcc@{\hspace{12pt}}cc@{\hspace{12pt}}cc@{}}
\toprule
& & & \multicolumn{2}{c}{$n=50{,}000$} & \multicolumn{2}{c}{$n=150{,}000$} & \multicolumn{2}{c}{$n=300{,}000$}\\
\cmidrule(lr){4-5}\cmidrule(lr){6-7}\cmidrule(lr){8-9}
$K$ & $d$ & $\lambda$ & Cov. & Length & Cov. & Length & Cov. & Length\\
\midrule
2 & 10 & 0 & 0.956 & 1.701 & 0.946 & 0.948 & 0.943 & 0.664\\
 &  & 0.1 & 0.952 & 1.664 & 0.955 & 0.925 & 0.958 & 0.648\\
 &  & 0.5 & 0.952 & 1.709 & 0.943 & 0.956 & 0.950 & 0.672\\
 &  & 1 & 0.948 & 1.821 & 0.945 & 1.028 & 0.944 & 0.724\\
 &  & 4 & 0.946 & 2.193 & 0.941 & 1.257 & 0.939 & 0.890\\
\addlinespace[2pt]
5 & 10 & 0 & 0.949 & 1.745 & 0.951 & 0.972 & 0.956 & 0.681\\
 &  & 0.1 & 0.954 & 1.735 & 0.961 & 0.965 & 0.953 & 0.677\\
 &  & 0.5 & 0.954 & 1.727 & 0.953 & 0.962 & 0.955 & 0.675\\
 &  & 1 & 0.950 & 1.755 & 0.945 & 0.982 & 0.953 & 0.689\\
 &  & 4 & 0.941 & 2.020 & 0.947 & 1.153 & 0.949 & 0.815\\
\bottomrule
\end{tabular}
\end{table}

\begin{table}[!htbp]
\spacingset{1.0}
\centering
\caption{Arm-3 augmented 95\% interval calibration, $d=10$: smoothed SVM loss. Entries are Monte Carlo mean coordinatewise coverage and mean interval length ($\times100$) across 200 streams.}
\label{tab:simulation-asymptotic-smooth_svm}
\small
\renewcommand{\arraystretch}{1.00}
\setlength{\tabcolsep}{3pt}
\begin{tabular}{@{}rrrcc@{\hspace{12pt}}cc@{\hspace{12pt}}cc@{}}
\toprule
& & & \multicolumn{2}{c}{$n=50{,}000$} & \multicolumn{2}{c}{$n=150{,}000$} & \multicolumn{2}{c}{$n=300{,}000$}\\
\cmidrule(lr){4-5}\cmidrule(lr){6-7}\cmidrule(lr){8-9}
$K$ & $d$ & $\lambda$ & Cov. & Length & Cov. & Length & Cov. & Length\\
\midrule
2 & 10 & 0 & 0.947 & 2.095 & 0.943 & 1.165 & 0.946 & 0.816\\
 &  & 0.1 & 0.944 & 2.052 & 0.946 & 1.140 & 0.947 & 0.799\\
 &  & 0.5 & 0.944 & 2.030 & 0.948 & 1.129 & 0.950 & 0.791\\
 &  & 1 & 0.957 & 2.116 & 0.950 & 1.181 & 0.949 & 0.829\\
 &  & 4 & 0.949 & 3.456 & 0.946 & 1.987 & 0.949 & 1.409\\
\addlinespace[2pt]
5 & 10 & 0 & 0.950 & 2.199 & 0.949 & 1.224 & 0.949 & 0.857\\
 &  & 0.1 & 0.951 & 2.194 & 0.949 & 1.220 & 0.953 & 0.855\\
 &  & 0.5 & 0.949 & 2.184 & 0.953 & 1.215 & 0.952 & 0.851\\
 &  & 1 & 0.947 & 2.197 & 0.950 & 1.223 & 0.951 & 0.858\\
 &  & 4 & 0.948 & 2.463 & 0.945 & 1.387 & 0.953 & 0.976\\
\bottomrule
\end{tabular}
\end{table}

\begin{table}[!htbp]
\spacingset{1.0}
\centering
\caption{Arm-3 augmented 95\% interval calibration, $d=10$: linear-regression squared loss. Entries are Monte Carlo mean coordinatewise coverage and mean interval length ($\times100$) across 200 streams.}
\label{tab:simulation-asymptotic-squared}
\small
\renewcommand{\arraystretch}{1.00}
\setlength{\tabcolsep}{3pt}
\begin{tabular}{@{}rrrcc@{\hspace{12pt}}cc@{\hspace{12pt}}cc@{}}
\toprule
& & & \multicolumn{2}{c}{$n=50{,}000$} & \multicolumn{2}{c}{$n=150{,}000$} & \multicolumn{2}{c}{$n=300{,}000$}\\
\cmidrule(lr){4-5}\cmidrule(lr){6-7}\cmidrule(lr){8-9}
$K$ & $d$ & $\lambda$ & Cov. & Length & Cov. & Length & Cov. & Length\\
\midrule
2 & 10 & 0 & 0.955 & 1.826 & 0.952 & 1.014 & 0.955 & 0.710\\
 &  & 0.1 & 0.950 & 1.829 & 0.946 & 1.015 & 0.946 & 0.710\\
 &  & 0.5 & 0.955 & 1.927 & 0.944 & 1.073 & 0.941 & 0.752\\
 &  & 1 & 0.952 & 2.159 & 0.949 & 1.214 & 0.950 & 0.854\\
 &  & 4 & 0.956 & 3.592 & 0.940 & 2.075 & 0.948 & 1.471\\
\addlinespace[2pt]
5 & 10 & 0 & 0.950 & 1.714 & 0.953 & 0.952 & 0.953 & 0.667\\
 &  & 0.1 & 0.945 & 1.712 & 0.952 & 0.951 & 0.950 & 0.666\\
 &  & 0.5 & 0.951 & 1.730 & 0.946 & 0.960 & 0.948 & 0.672\\
 &  & 1 & 0.949 & 1.787 & 0.946 & 0.996 & 0.942 & 0.698\\
 &  & 4 & 0.951 & 2.478 & 0.953 & 1.425 & 0.954 & 1.009\\
\bottomrule
\end{tabular}
\end{table}

\FloatBarrier

\begin{figure}[!htbp]
\centering
\includegraphics[width=\textwidth]{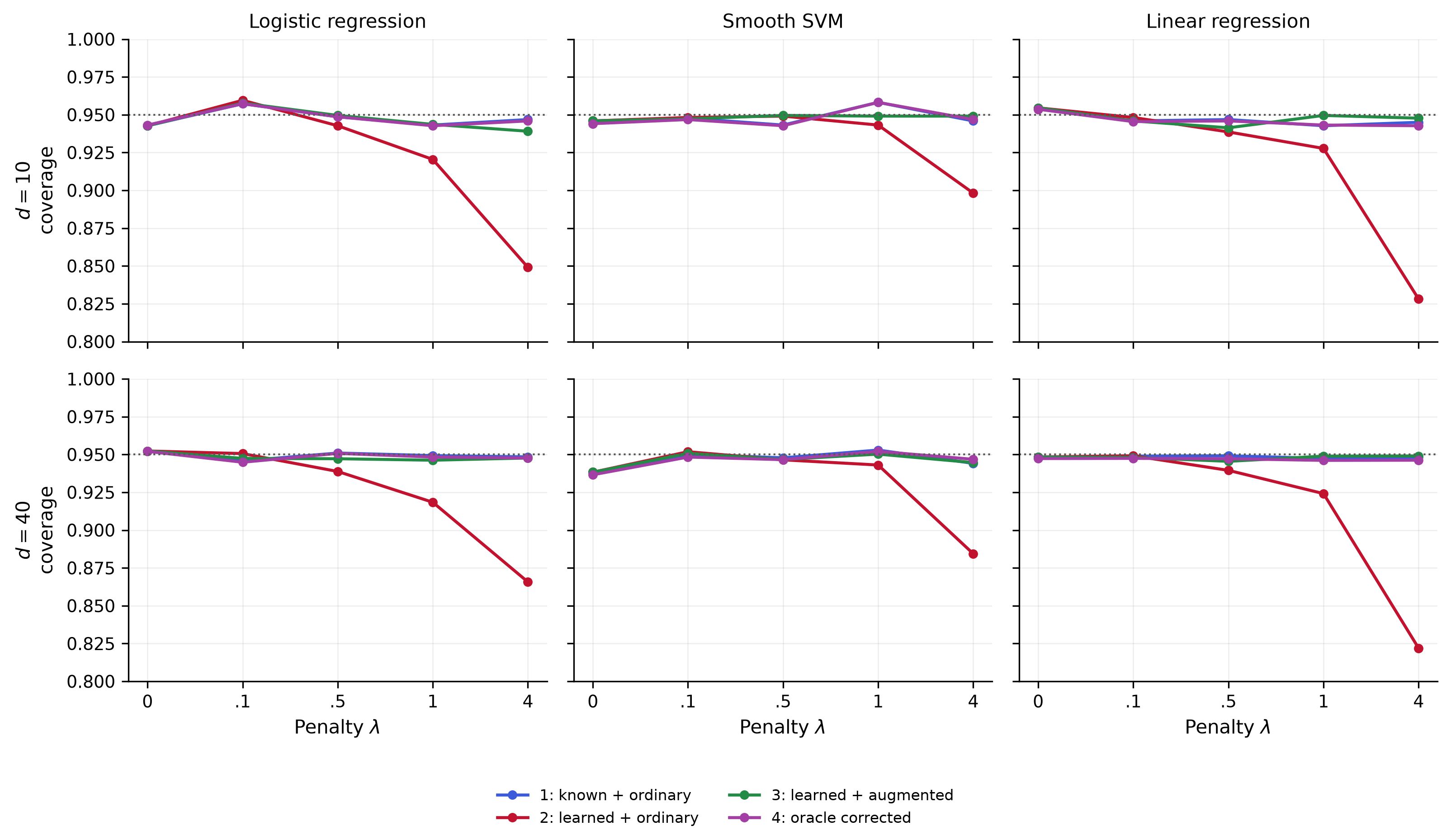}
\caption{Four-arm mean coordinatewise coverage at \(n=300{,}000\) for
\(K=2\), \(d\in\{10,40\}\), and the three losses; the dotted line marks
nominal 95\% coverage.}
\label{fig:four-arm-coverage}
\end{figure}
\FloatBarrier

Figure \ref{fig:four-arm-coverage} shows that Arm 2 undercovers as
\(\lambda\) increases, whereas Arms 1, 3, and 4 remain close to nominal
95\% coverage. Arm 1 is
calibrated when the operator is fixed. Arm 2 departs from nominal coverage as
\(\lambda\) grows despite sharing its center with Arm 3, so the defect is not
optimization error. Arm 3 recovers nominal coverage and agrees with the
infeasible path-adjusted Arm 4. This separation is strongest at large
\(\lambda\), exactly where the harmonic operator term enters the first-order
law with the greatest weight. Stability across the three losses and both
dimensions supports the augmented covariance formula. The supplement reports
both group counts, all checkpoints, interval-length comparisons, and
fairness--utility summaries.

\section{Real-data applications}\label{sec:real-data}
\renewcommand{\theequation}{6.\arabic{equation}}\setcounter{equation}{0}

The applications illustrate target movement and streaming precision. They do
not provide repeated-stream coverage evidence; that role belongs to Section
\ref{sec4}. Fairness summaries are descriptive and have no causal
interpretation.

\subsection{Adult income classification}

We classify whether Adult income exceeds \$50,000 using 36,177 training and
9,045 test observations. Training-standardized numeric variables and one-hot
encoded factors give 102 coordinates including the intercept; sex defines the
two groups but is excluded from the predictors. Logistic and smooth-hinge
losses use the raw-covariance score penalty at
\(\lambda\in\{0,0.1,0.25,0.5,1,2\}\). Held-out utility is accuracy and
fairness is the absolute group difference in positive prediction rates. The
supplement gives preprocessing, tuning, target-distance, and interval-length
details for the full grid.

Between \(\lambda=0\) and 2, the gap falls by 41\% for both losses, while
accuracy decreases by 1.2 percentage points for logistic regression and 0.9
for the smooth SVM.

\begin{figure}[!t]
\centering
\includegraphics[width=.80\textwidth]{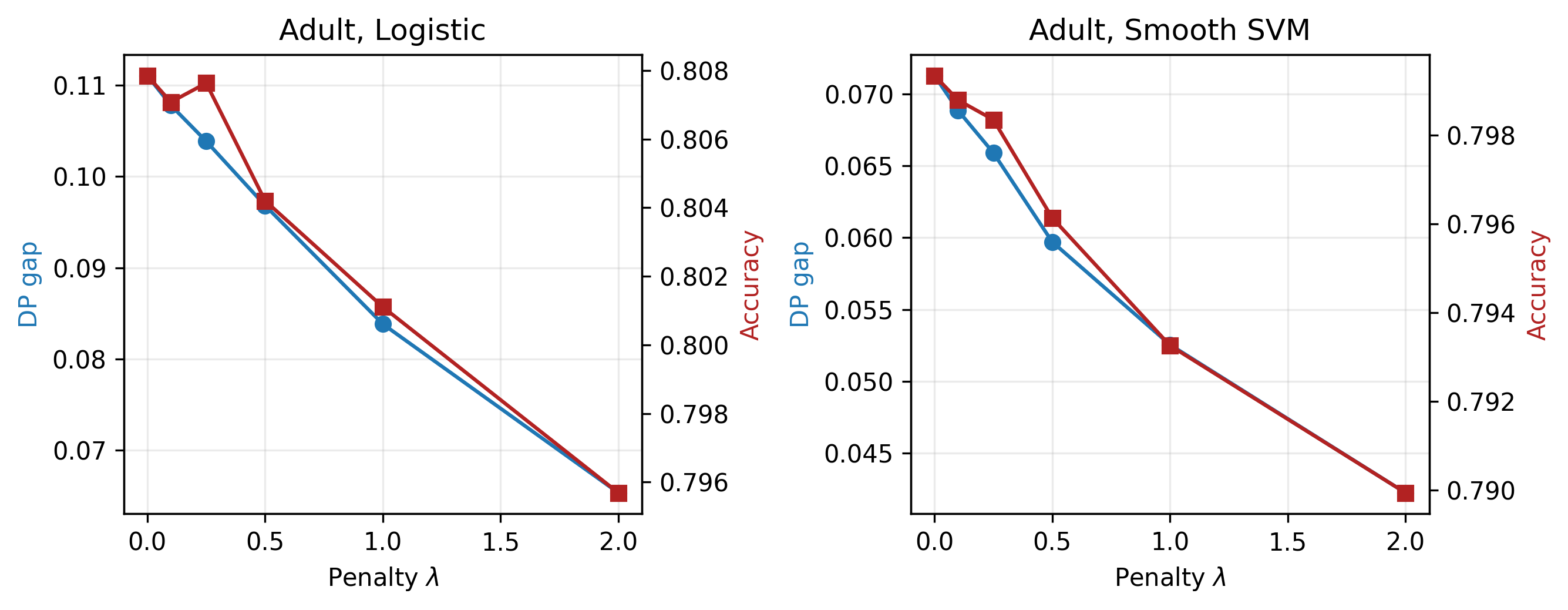}
\caption{Adult fairness--utility trade-off over the full penalty grid.
Blue circles give the held-out hard-decision demographic-parity gap, and red
squares give held-out accuracy, for logistic regression and the smoothed SVM.}
\label{fig:adult-tradeoff-main}
\end{figure}

Figure \ref{fig:adult-tradeoff-main} shows that the disparity reduction is
progressive, with most of the improvement occurring before the largest
penalty and comparatively slow changes in held-out accuracy. Interval lengths
remain nearly flat because the operator correction and curvature change
largely offset. The curves describe candidate operating points rather than a
data-driven choice of \(\lambda\); the complete grid is in the supplement.

The logistic response is directional rather than uniform shrinkage. From
\(\lambda=0\) to 2, hours per week changes most among the five displayed
continuous covariates, falling by 0.053; age, capital gain, and capital loss
fall by 0.009--0.013, while education number rises slightly. These conditional
movements reflect the group-correlated direction
\(\boldsymbol C\boldsymbol\beta\), not variable-level fairness scores.

\begin{figure}[!t]
\centering
\includegraphics[width=.66\textwidth]{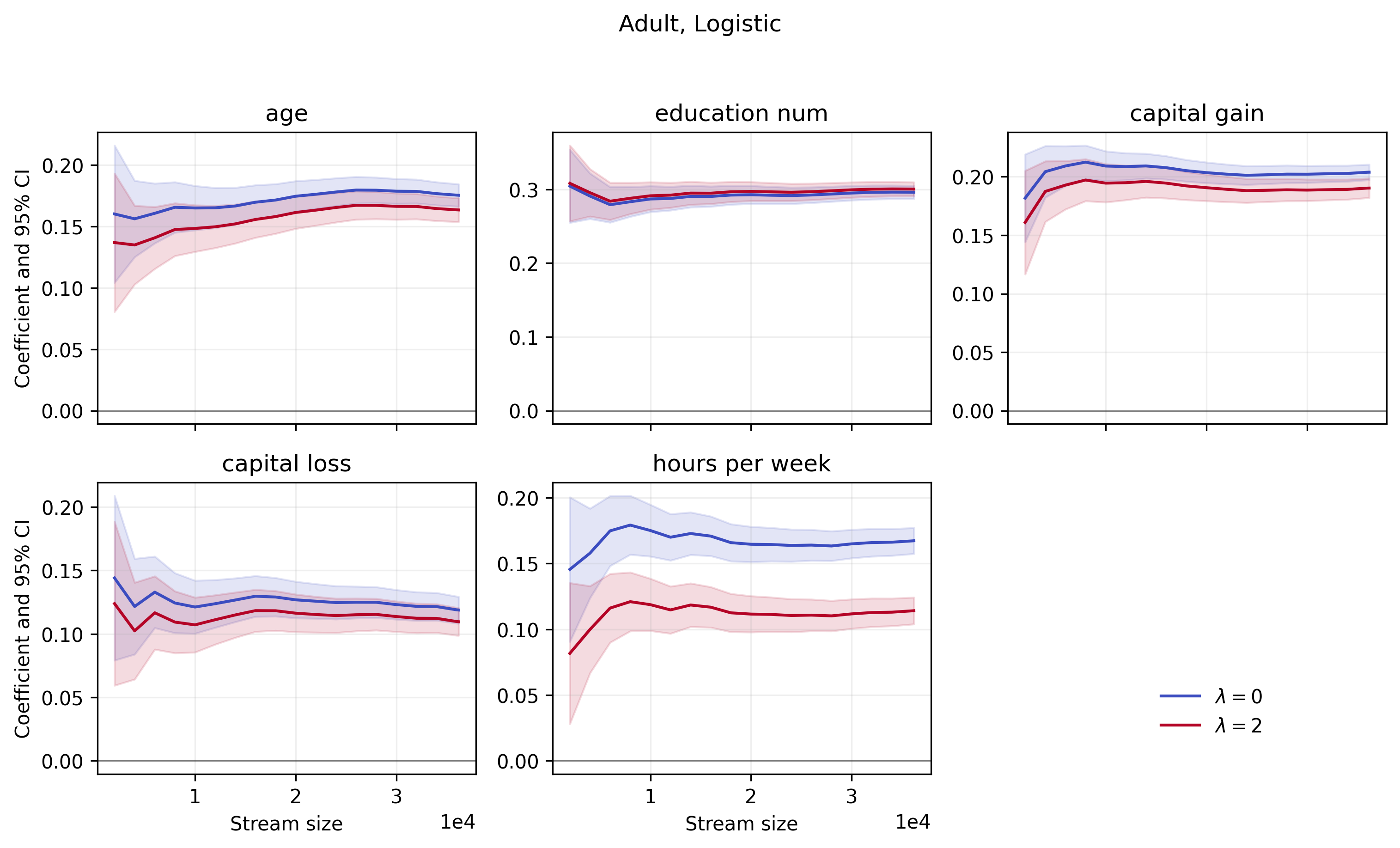}
\caption{Adult logistic-regression coefficient trajectories with pointwise
95\% augmented confidence bands for $\lambda=0$ and 2. The five subpanels
display the continuous coordinates used to summarize penalty-induced target
movement. Persistent terminal separation is estimand movement, whereas
contracting bands show decreasing streaming uncertainty.}
\label{fig:adult-ci-main}
\end{figure}

Figure \ref{fig:adult-ci-main} separates early optimization noise from target
movement: the bands contract with stream size, but the terminal
hours-per-week paths remain separated, whereas education number is relatively
insensitive. Full six-penalty shifts and smooth-SVM trajectories are in the
supplement.

\subsection{IPUMS wage regression}

We analyze IPUMS log real hourly wages separately for 1980, 1990, 2000,
2001--2005, 2006--2010, and 2011--2015 among White wage/salary workers aged
18--65 who work at least 30 hours weekly and satisfy the source restrictions.
The common 53-coordinate design contains gender, education, age band, Census
division, and prespecified interactions. Gender and age remain predictors;
separate codes define gender or four-group age parity. Each period is streamed
once with linear regression at \(\lambda\in\{0,1,4,10,20\}\); full selection,
coding, and tuning details are in the supplement. With the identity link, the
decision score is the prediction, so fairness is
\begin{equation}\label{eq:ipums-dp-gap}
 \widehat\Delta_{\rm DP}(\boldsymbol\beta)
 =\max_g\widehat\E\{\boldsymbol\beta^\top\boldsymbol X\mid Z=g\}
 -\min_g\widehat\E\{\boldsymbol\beta^\top\boldsymbol X\mid Z=g\}.
\end{equation}
Here \(\widehat\E\) denotes the empirical mean over the observations in the
indicated group.
This targets equality of expected predictions, not subgroup MSE; MSE remains
the utility measure.

The combined fairness--utility paths show a
monotone decline in the predicted-mean gap for both targets. At
\(\lambda=4\), the combined gender gap falls by 66.1\% with a 2.75\% MSE
increase, while the combined age gap falls by 47.5\% with a 2.85\% MSE
increase. At \(\lambda=20\), the reductions are 90.7\% and 81.7\%, paired
with MSE increases of 5.18\% and 8.28\%. These numbers describe operating
points on a fitted frontier; the inferential theory does not select a penalty.

\begin{figure}[!t]
\centering
\includegraphics[width=.86\textwidth]{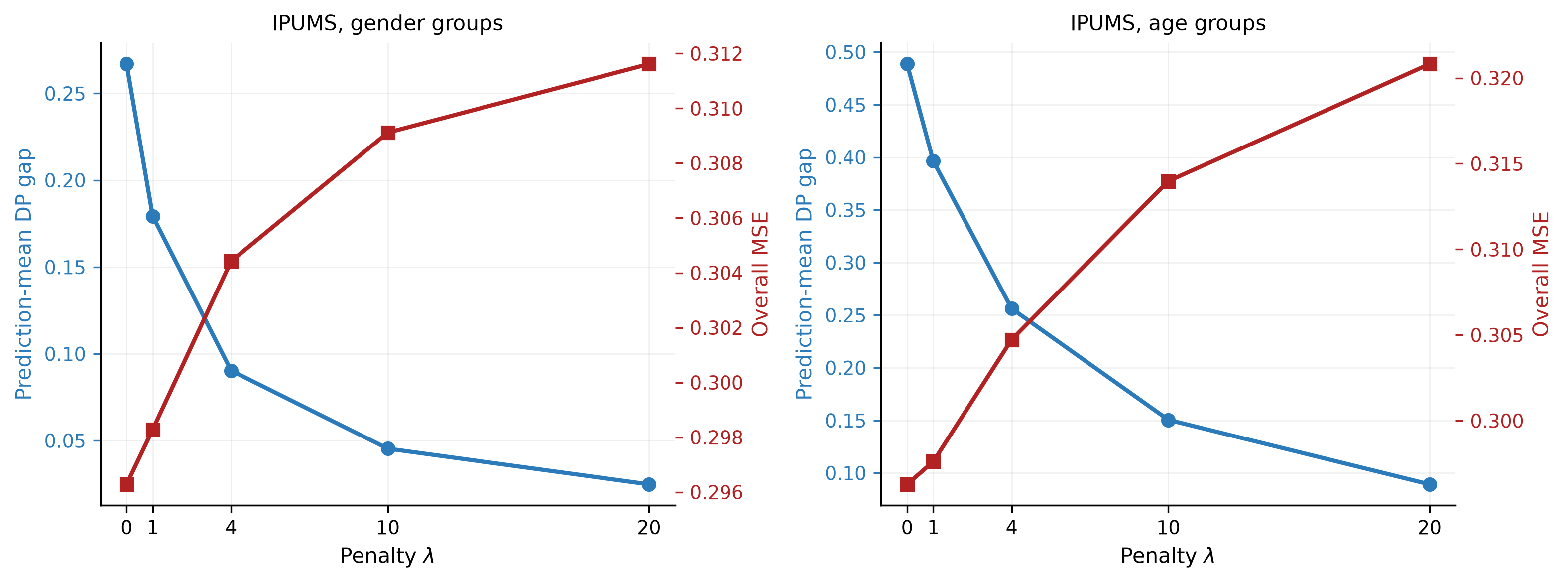}
\caption{Combined IPUMS fairness--utility trade-off. Blue circles give the
sample-size-weighted predicted-mean gap and red squares the weighted MSE
across the six separate-period fits. The panels impose gender parity and
four-group age parity, respectively; the displayed summaries are not pooled
refits.}
\label{fig:ipums-tradeoff-main}
\end{figure}

Figure \ref{fig:ipums-tradeoff-main} makes the curvature of the two frontiers
visible. Gender parity achieves a sharp initial gap reduction with a modest
MSE change, whereas the age-parity frontier is more gradual and becomes
costlier at the largest penalties.

Every period has a smaller gap at \(\lambda=4\), although early cohorts pay the
largest utility cost. Female changes most in all six gender fits, with weighted
absolute endpoint shift 0.240 from \(\lambda=0\) to 20. For age parity,
Age 40--49 leads in 1980 and 1990 and Age 50--65 thereafter; their combined
absolute endpoint shifts are 0.350 and 0.351. Complete period-specific
rankings are in the supplement.

The period-specific paths distinguish a stable fairness response from a
changing utility cost. At \(\lambda=4\), the gender-gap reduction is tightly
concentrated between 65.4\% and 66.6\% across all six periods, but the MSE
increase falls from 8.40\% in 1980 to 1.25\% in 2011--2015. The age-gap
reduction is larger in 1980--1990 (52.7\% and 51.0\%) than after 2000
(45.3\%--46.8\%), while its MSE cost remains between 2.30\% and 2.78\% after
2000. Thus a similar penalty can purchase a different utility--fairness
trade-off as the wage structure and covariate distribution change.

\begin{figure}[!t]
\centering
\includegraphics[width=.80\textwidth]{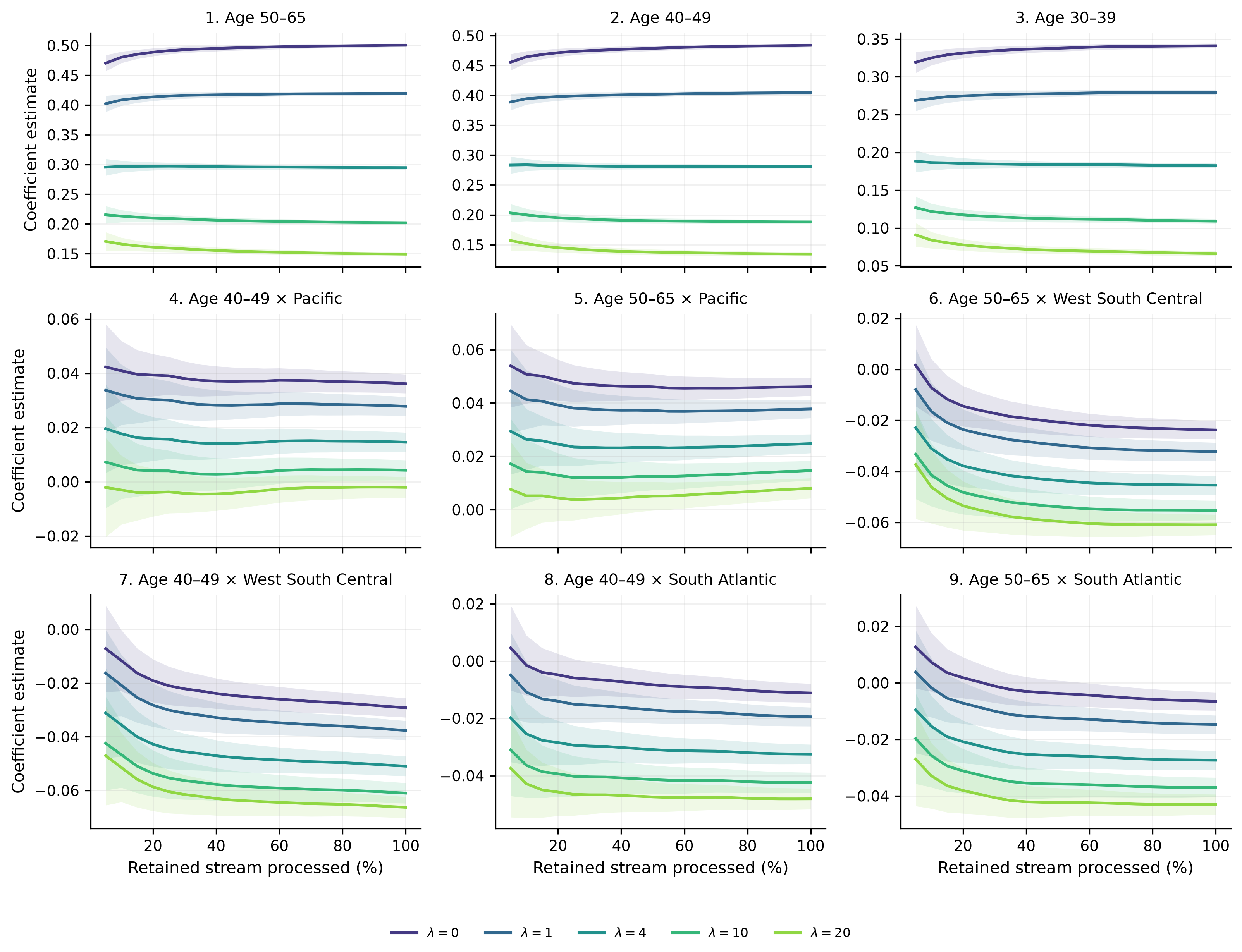}
\caption{The nine coordinates most changed by the IPUMS age-parity penalty,
ranked by the sample-size-weighted absolute endpoint shift from $\lambda=0$
to 20 across the six periods. Each panel shows the corresponding
sample-size-weighted coefficient trajectory and pointwise 95\% augmented
bands for all five penalties. These are weighted summaries of separate-period
fits, not a pooled refit, and the bands are pointwise in stream size.}
\label{fig:ipums-confidence-paths-main}
\end{figure}

The coordinate rankings show the same temporal heterogeneity. Under gender
parity, Female remains the dominant direction, but its absolute endpoint
shift decreases steadily from 0.375 in 1980 to 0.176 in 2011--2015. Under age
parity, the largest adjustment moves from Age 40--49 in the first two periods
to Age 50--65 thereafter, and the latter shift reaches 0.405 in 2011--2015.
The sample-size-weighted trajectories summarize these consistent directions,
but they do not replace the separate-period fits: a pooled coefficient would
mix period composition with within-period penalty response.

The changes are smooth, not common shrinkage. For age parity, weighted Age
30--39, Age 40--49, and Age 50--65 coefficients fall from 0.341, 0.484, and
0.500 at \(\lambda=0\) to 0.066, 0.134, and 0.149 at \(\lambda=20\). The next
six coordinates are Age 40--49 or Age 50--65 interactions with Pacific, West
South Central, or South Atlantic; their weighted coefficient changes are
negative, with absolute shifts ranging from 0.036 to 0.038. The supplement
reports the complete gender path,
including the Female coefficient's sign reversal as \(\lambda\) increases
from 1 to 4.

Figure \ref{fig:ipums-confidence-paths-main} explains the ranking. The three
age indicators enter group-mean score contrasts directly and have the largest
fairness-gradient loadings, so all move sharply downward as \(\lambda\)
increases. Six age-by-division terms follow because they redistribute regional
score components correlated with those contrasts; they also decline, but less
than the main effects. Persistent terminal separation is target movement, not
early-stream noise. The ranking is conditional and descriptive, not causal;
gender and period-specific paths remain in the supplement.

\FloatBarrier
\section{Discussion}\label{sec:conclusion}

We developed fixed-memory inference for prediction coefficients when a
score-covariance penalty is estimated during training. Repeated use of the
same observations creates a harmonic influence term, which the augmented
covariance accounts for while retaining the Polyak average. Simulations and
Adult and IPUMS applications illustrate calibration, coefficient movement,
and fairness--utility trade-offs. Future work could extend inference to
dependent streams, adaptive penalties, and learned representations, and study
conditions linking score parity to threshold-decision parity.

\bibliography{reference}

@article{schifano2016online,
  title={Online updating of statistical inference in the big data setting},
  author={Schifano, Elizabeth D and Wu, Jing and Wang, Chun and Yan, Jun and Chen, Ming-Hui},
  journal={Technometrics},
  volume={58},
  number={3},
  pages={393--403},
  year={2016},
  publisher={Taylor \& Francis},
  doi={10.1080/00401706.2016.1142900},
  url={https://www.tandfonline.com/doi/abs/10.1080/00401706.2016.1142900}
}

@article{lin2011aggregated,
  title={Aggregated estimating equation estimation},
  author={Lin, Nan and Xi, Ruibin},
  journal={Statistics and its Interface},
  volume={4},
  number={1},
  pages={73--83},
  year={2011},
  publisher={International Press},
  doi={10.4310/SII.2011.v4.n1.a8},
  url={https://intlpress.com/site/pub/files/_fulltext/journals/sii/2011/0004/0001/SII-2011-0004-0001-a008.pdf}
}

@article{robbins1951stochastic,
  title={A stochastic approximation method},
  author={Robbins, Herbert and Monro, Sutton},
  journal={The Annals of Mathematical Statistics},
  volume={22},
  number={3},
  pages={400--407},
  year={1951},
  publisher={Institute of Mathematical Statistics},
  doi={10.1214/aoms/1177729586},
  url={https://projecteuclid.org/journals/annals-of-mathematical-statistics/volume-22/issue-3/A-Stochastic-Approximation-Method/10.1214/aoms/1177729586.full}
}

@article{polyak1992acceleration,
  title={Acceleration of stochastic approximation by averaging},
  author={Polyak, Boris T and Juditsky, Anatoli B},
  journal={SIAM Journal on Control and Optimization},
  volume={30},
  number={4},
  pages={838--855},
  year={1992},
  publisher={SIAM},
  doi={10.1137/0330046},
  url={https://epubs.siam.org/doi/10.1137/0330046}
}

@article{chen2020statistical,
  title={Statistical inference for model parameters in stochastic gradient descent},
  author={Chen, Xi and Lee, Jason D and Tong, Xin T and Zhang, Yichen},
  journal={The Annals of Statistics},
  volume={48},
  number={1},
  pages={251--273},
  year={2020},
  doi={10.1214/18-AOS1801},
  url={https://projecteuclid.org/journals/annals-of-statistics/volume-48/issue-1/Statistical-inference-for-model-parameters-in-stochastic-gradient-descent/10.1214/18-AOS1801.full}
}

@article{fang2018online,
  title={Online Bootstrap Confidence Intervals for the Stochastic Gradient Descent Estimator},
  author={Fang, Yixin and Xu, Jinfeng and Yang, Lei},
  journal={Journal of Machine Learning Research},
  volume={19}, number={78}, pages={1--21}, year={2018},
  url={https://www.jmlr.org/papers/v19/17-370.html}
}

@inproceedings{liu2024empirical,
  title={Empirical Likelihood for Fair Classification},
  author={Liu, Pangpang and Zhao, Yichuan},
  booktitle={International Conference on Learning Representations},
  year={2024},
  url={https://proceedings.iclr.cc/paper_files/paper/2024/hash/39fac857b4467e3ef4f358186bb07d81-Abstract-Conference.html}
}

@inproceedings{roy2026fairness,
  title={Fairness Uncertainty Quantification: A Constrained Stochastic Optimization Perspective},
  author={Roy, Abhishek and Mohapatra, Prasant},
  booktitle={Proceedings of the 42nd Conference on Uncertainty in Artificial Intelligence},
  series={Proceedings of Machine Learning Research},
  volume={337}, pages={5845--5865}, year={2026}, publisher={PMLR},
  url={https://proceedings.mlr.press/v337/roy26b.html}
}

@article{zhu2023online,
  title={Online covariance matrix estimation in stochastic gradient descent},
  author={Zhu, Wanrong and Chen, Xi and Wu, Wei Biao},
  journal={Journal of the American Statistical Association},
  volume={118},
  number={541},
  pages={393--404},
  year={2023},
  publisher={Taylor \& Francis},
  doi={10.1080/01621459.2021.1933498},
  url={https://www.tandfonline.com/doi/abs/10.1080/01621459.2021.1933498}
}

@article{han2024online,
  title={Online inference with debiased stochastic gradient descent},
  author={Han, Ruijian and Luo, Lan and Lin, Yuanyuan and Huang, Jian},
  journal={Biometrika},
  volume={111},
  number={1},
  pages={93--108},
  year={2024},
  publisher={Oxford University Press},
  doi={10.1093/biomet/asad046},
  url={https://academic.oup.com/biomet/article/111/1/93/7232226}
}

@inproceedings{dwork2012fairness,
  title={Fairness through awareness},
  author={Dwork, Cynthia and Hardt, Moritz and Pitassi, Toniann and Reingold, Omer and Zemel, Richard},
  booktitle={Proceedings of the 3rd Innovations in Theoretical Computer Science Conference},
  pages={214--226},
  year={2012},
  publisher={Association for Computing Machinery},
  doi={10.1145/2090236.2090255},
  url={https://dl.acm.org/doi/10.1145/2090236.2090255}
}

@article{hardt2016equality,
  title={Equality of opportunity in supervised learning},
  author={Hardt, Moritz and Price, Eric and Srebro, Nati},
  journal={Advances in Neural Information Processing Systems},
  volume={29},
  pages={3315--3323},
  year={2016},
  url={https://proceedings.neurips.cc/paper_files/paper/2016/hash/6a9659feb1216f14f7384ba499518b38-Abstract.html}
}

@book{barocas2023fairness,
  title={Fairness and Machine Learning: Limitations and Opportunities},
  author={Barocas, Solon and Hardt, Moritz and Narayanan, Arvind},
  year={2023},
  publisher={MIT Press},
  isbn={9780262048613},
  url={https://mitpress.mit.edu/9780262048613/fairness-and-machine-learning/}
}

@inproceedings{kamishima2012fairness,
  title={Fairness-aware classifier with prejudice remover regularizer},
  author={Kamishima, Toshihiro and Akaho, Shotaro and Asoh, Hideki and Sakuma, Jun},
  booktitle={Machine Learning and Knowledge Discovery in Databases},
  pages={35--50},
  year={2012},
  publisher={Springer},
  doi={10.1007/978-3-642-33486-3_3},
  url={https://link.springer.com/chapter/10.1007/978-3-642-33486-3_3}
}

@article{zafar2019fairness,
  title={Fairness constraints: A flexible approach for fair classification},
  author={Zafar, Muhammad Bilal and Valera, Isabel and Gomez-Rodriguez, Manuel and Gummadi, Krishna P},
  journal={Journal of Machine Learning Research},
  volume={20},
  number={75},
  pages={1--42},
  year={2019},
  url={https://jmlr.org/papers/v20/18-262.html}
}

@inproceedings{agarwal2019fair,
  title={Fair regression: Quantitative definitions and reduction-based algorithms},
  author={Agarwal, Alekh and Dud{\'i}k, Miroslav and Wu, Zhiwei Steven},
  booktitle={Proceedings of the 36th International Conference on Machine Learning},
  pages={120--129},
  year={2019},
  volume={97},
  series={Proceedings of Machine Learning Research},
  url={https://proceedings.mlr.press/v97/agarwal19d.html}
}

@article{caton2024fairness,
  title={Fairness in machine learning: A survey},
  author={Caton, Simon and Haas, Christian},
  journal={ACM Computing Surveys},
  volume={56},
  number={7},
  pages={1--38},
  year={2024},
  publisher={Association for Computing Machinery},
  doi={10.1145/3616865},
  url={https://dl.acm.org/doi/10.1145/3616865}
}

@inproceedings{zhang2018mitigating,
  title={Mitigating unwanted biases with adversarial learning},
  author={Zhang, Brian Hu and Lemoine, Blake and Mitchell, Margaret},
  booktitle={Proceedings of the 2018 AAAI/ACM Conference on AI, Ethics, and Society},
  pages={335--340},
  year={2018},
  publisher={Association for Computing Machinery},
  doi={10.1145/3278721.3278779},
  url={https://dl.acm.org/doi/10.1145/3278721.3278779}
}

@inproceedings{dehdashtian2024ufate,
  title={Utility-fairness trade-offs and how to find them},
  author={Dehdashtian, Sepehr and Sadeghi, Bashir and Boddeti, Vishnu Naresh},
  booktitle={Proceedings of the IEEE/CVF Conference on Computer Vision and Pattern Recognition},
  pages={12037--12046},
  year={2024},
  month={June},
  url={https://openaccess.thecvf.com/content/CVPR2024/html/Dehdashtian_Utility-Fairness_Trade-Offs_and_How_to_Find_Them_CVPR_2024_paper.html}
}

@inproceedings{liu2018delayed,
  title={Delayed impact of fair machine learning},
  author={Liu, Lydia T and Dean, Sarah and Rolf, Esther and Simchowitz, Max and Hardt, Moritz},
  booktitle={Proceedings of the 35th International Conference on Machine Learning},
  pages={3150--3158},
  year={2018},
  volume={80},
  series={Proceedings of Machine Learning Research},
  publisher={\mbox{PMLR}},
  url={https://proceedings.mlr.press/v80/liu18c.html}
}

@article{fuster2022predictably,
  title={Predictably unequal? The effects of machine learning on credit markets},
  author={Fuster, Andreas and Goldsmith-Pinkham, Paul and Ramadorai, Tarun and Walther, Ansgar},
  journal={The Journal of Finance},
  volume={77},
  number={1},
  pages={5--47},
  year={2022},
  publisher={Wiley},
  doi={10.1111/jofi.13090},
  url={https://onlinelibrary.wiley.com/doi/10.1111/jofi.13090}
}

@inproceedings{potash2015predictive,
  title={Predictive modeling for public health: Preventing childhood lead poisoning},
  author={Potash, Eric and Brew, Joe and Loewi, Alexander and Majumdar, Subhabrata and Reece, Andrew and Walsh, Joe and Rozier, Eric and Jorgenson, Emile and Mansour, Raed and Ghani, Rayid},
  booktitle={Proceedings of the 21st ACM SIGKDD International Conference on Knowledge Discovery and Data Mining},
  pages={2039--2047},
  year={2015},
  publisher={Association for Computing Machinery},
  doi={10.1145/2783258.2788629},
  url={https://dl.acm.org/doi/10.1145/2783258.2788629}
}

@article{wei2021optimized,
  title={Optimized Score Transformation for Consistent Fair Classification},
  author={Wei, Dennis and Ramamurthy, Karthikeyan Natesan and Calmon, Flavio P.},
  journal={Journal of Machine Learning Research},
  volume={22},
  number={258},
  pages={1--78},
  year={2021},
  url={https://jmlr.org/papers/v22/20-1143.html}
}

@inproceedings{gitiaux2021learning,
  title={Learning Smooth and Fair Representations},
  author={Gitiaux, Xavier and Rangwala, Huzefa},
  booktitle={Proceedings of the 24th International Conference on Artificial Intelligence and Statistics},
  volume={130},
  pages={253--261},
  year={2021},
  publisher={\mbox{PMLR}},
  url={https://proceedings.mlr.press/v130/gitiaux21a.html}
}

@article{zhao2022inherent,
  title={Inherent Tradeoffs in Learning Fair Representations},
  author={Zhao, Han and Gordon, Geoffrey J.},
  journal={Journal of Machine Learning Research},
  volume={23},
  number={57},
  pages={1--26},
  year={2022},
  url={https://jmlr.org/papers/v23/21-1427.html}
}

@misc{rajkumar2022identification,
  title={Identification and Inference with Min-over-max Estimators for the Measurement of Labor Market Fairness},
  author={Rajkumar, Karthik},
  year={2022},
  howpublished={arXiv:2207.13797},
  doi={10.48550/arXiv.2207.13797},
  url={https://arxiv.org/abs/2207.13797}
}

@inproceedings{henzinger2023monitoring,
  title={Monitoring Algorithmic Fairness},
  author={Henzinger, Thomas A. and Karimi, Mahyar and Kueffner, Konstantin and Mallik, Kaushik},
  booktitle={Computer Aided Verification},
  series={Lecture Notes in Computer Science},
  volume={13965},
  pages={358--382},
  year={2023},
  publisher={Springer},
  doi={10.1007/978-3-031-37703-7_17},
  url={https://link.springer.com/chapter/10.1007/978-3-031-37703-7_17}
}

@article{cherian2024statistical,
  title={Statistical Inference for Fairness Auditing},
  author={Cherian, John J. and Cand{\`e}s, Emmanuel J.},
  journal={Journal of Machine Learning Research},
  volume={25},
  number={149},
  pages={1--49},
  year={2024},
  url={https://jmlr.org/papers/v25/23-0739.html}
}

@article{denis2024fairness,
  title={Fairness Guarantees in Multi-Class Classification with Demographic Parity},
  author={Denis, Christophe and Elie, Romuald and Hebiri, Mohamed and Hu, Fran{\c{c}}ois},
  journal={Journal of Machine Learning Research},
  volume={25},
  number={130},
  pages={1--46},
  year={2024},
  url={https://jmlr.org/papers/v25/23-0322.html}
}
\bibliographystyle{chicago}
\end{document}